\documentclass[useAMS,usenatbib]{mn2e}
\usepackage{graphicx}
\usepackage{epsfig}

\title[Spectral Study of Type 1 AGN - II. Relation Between X-ray Emission and Optical Spectra]{A Combined Optical and X-ray Study of Unobscured Type 1 AGN. II. Relation Between X-ray Emission and Optical Spectra}

\author[C. Jin, M. Ward, C. Done]
{Chichuan Jin\thanks{E-mail: chichuan.jin@durham.ac.uk},
Martin Ward, Chris Done\\
Department of Physics, University of Durham, South Road, Durham, DH1 3LE, UK}

\begin{document}

\date{Submitted to MNRAS}

\maketitle

\label{firstpage}

\begin{abstract}
In this second paper in a series of three,
we study the properties of the various emission features
and underlying continuum in the optical spectra of Type 1 active
galactic nuclei (AGNs) by using the unobscured hard X-ray emission as
a diagnostic. We introduce the use of the `Correlation Spectrum
Technique' (CST) for the first time. We use this to show the strength
of correlation between the hard X-ray luminosity and each wavelength
of the optical spectrum.  This shows that for Broad Line Seyfert 1s
all the strong emission lines (broad component of H$\alpha$ and
H$\beta$, [NeIII] $\lambda\lambda$3869/3967, [OI]
$\lambda\lambda$6300/6364, [OII] $\lambda\lambda$3726/3729, [OIII]
$\lambda\lambda$4959/5007) and the optical underlying continuum all
strongly correlate with the hard X-ray emission. By contrast, the
Narrow Line Seyfert 1s show stronger correlation of the optical
continuum but weaker correlation in the lines.

A cross-correlation with luminosity between the various Balmer line
components and the broadband SED components shows that the best
correlation exists between the hard X-ray component and broad
component (BC) of the Balmer lines. Such a correlation is weaker for
the intermediate (IC) and narrow components, which supports the view
that broad line region (BLR) has the closest link with the AGN's
compact X-ray emission. The equivalent widths of Balmer line IC and BC
are found to correlate with L$_{2-10keV}$, $\kappa_{2-10keV}^{-1} =
L_{bol}/L_{2-10keV}$, Balmer line FWHM and black hole mass.  There is
a non-linear dependence of the Balmer line IC and BC luminosities
with L$_{2-10keV}$ and L$_{5100}$, which suggests that a second-order
factor such as the ILR and BLR covering factors affect the Balmer line
component luminosities.  The Balmer decrement is found to decrease
from $\sim$5 in the line core to $\sim$2 in the extended wings, with
mean decrements of 2.1 in BLR and 4.8 in ILR.  This suggests different
physical conditions in these regions, such as variations in electron
density and dust abundance.

The [OIII] line is composed of a narrow core together with a
blue-shifted component with average outflow velocity of
$130^{+230}_{-80}~km~s^{-1}$.  The total luminosity of [OIII]
$\lambda$5007 shows the best correlation with the luminosity of hard
X-ray emission, and so can be used to estimate the intrinsic X-ray
luminosity of obscured AGNs. We use the CST to show the correlation of
the [OIII] $\lambda$5007 luminosity with each wavelength of the full
continuum SED. This shows as before that the [OIII] $\lambda$5007
strongly correlates with power law tail, but also that it correlates
almost as strongly with the optical continuum from the disc, but not
with the soft excess.

\end{abstract}


\section{Introduction}
\label{section:intro}
Active Galactic Nuclei (AGNs) are characterized by strong energy
output over a very broad frequency range, from the infrared to hard
X-ray.  Although their spectral energy distribution (SED) has been
studied for several decades, one of the main difficulties is the
unobservable region between 0.01 keV and 0.2 keV, which is due to
Galactic and intrinsic photoelectric absorption.  As a result most
previous studies have focused mainly either on the UV/optical/infrared
side or the soft/hard X-ray side, of the unobservable region.

\subsection{Previous Work}
In the optical band, information about the AGN can be derived from 
the Balmer emission line series e.g. H$\alpha$, 
H$\beta$ and H$\gamma$, strong forbidden lines e.g. the  
[OIII]$\lambda$4959/5007 doublets, and the optical underlying
continuum. Timing analysis of variations in the Balmer lines
via Reverberation Mapping study 
is the most reliable method to estimate AGN black hole 
masses (e.g. Kaspi et al. 2000; Peterson et al. 2004).
Measurement of the [OIII] $\lambda$5007 emission line is used to 
provide an orientation independent estimate of the AGN's 
central ionizing flux (Kauffmann et al. 2003; Heckman et al. 2004;
Brinchmann et al. 2004; Heckman et al. 2005). It is also a principal 
diagnostic of the properties of the narrow line region (NLR),
such as the stellar velocity dispersion (e.g.
Nelson \& Whittle 1995; Nelson \& Whittle 1996; Boroson 2003; Greene \& Ho 2005)
and outflow speed (e.g. Bian, Yuan \& Zhao 2005; Komossa et al. 2008).
The underlying continuum in the optical is believed to be  
dominated by emission from the accretion disc (Koratkar \& Blaes 1999;
Jin et al. 2011, hereafter: Paper-I) and AGN's host galaxy
(e.g. Hao et al. 2010; Landt et al. 2011).

The soft to hard X-ray emission is emitted from a compact region.  The
hard X-ray power law tail above 2 keV is produced by disc photons
which are Compton up-scattered by a hot, optically thin electron
population (e.g. Haardt \& Maraschi 1991; Zdziarski, Poutanen \&
Johnson 2000).  However, the origin of the soft X-ray excess below 1
keV is still not clear (e.g. Gierli\'{n}ski \& Done 2004; Done et
al. 2011; also see Paper-I).  Significant absorption of the continuum
by ionized gas located near to the central region (i.e. the `Warm
Absorber') can further complicate interpretation of the observed X-ray
spectra (Reynolds 1997; Crenshaw \& Kraemer 1999).

The infrared band especially beyond 2 $\mu$m is believed to be dominated by emission
from the dusty torus, and so provides information about its physical 
condition (e.g. Rees et al. 1969; Rieke 1978; Lebofsky \& Reike 1980).
The observed 1 $\mu$m minimum in the SED
is also a common feature (Sanders et al. 1989).
Recently, Landt et al. (2011) showed that the continuum luminosity at 1 $\mu$m can be
used to estimate the AGN black hole mass.

Many previous studies have attempted to construct AGN broadband SED 
by combining multi-waveband observational data to model the emission across the
UV-soft X-ray energy gap (see references in Paper-I).
Whichever method is used to construct the broadband SED, it
is found that in almost all cases the SED peaks in the unobservable 
region (rest-frame) which encompasses the `Big Blue Bump (BBB)'. 
One important consequence of this is that the unobservable region 
contains the largest fraction of an AGN's radiated energy
(Walter \& Fink 1993; Grupe et al. 1998, 1999; Paper-I).

\subsection{The Approach Adopted in This Paper}
In Paper-I, we presented modelling results for a sample of
51 unobscured nearby Type 1 AGNs.
From the spectral fitting we derived numerous spectral parameters that were not 
contaminated by X-ray or optical obscuration.
{\bf In this paper, we investigate the link between the central ionization flux
characterized by the unobscured hard X-ray, and the properties of the various optical
emission lines and the continuum. We conduct this investigation by studying the profiles
of various emission lines, and the luminosity correlations between the hard X-ray
continuum and various optical emission features.}

Narrow Line Seyfert 1s (NLS1s) are often considered as a special type of
AGN whose permitted line width is comparable to other forbidden lines,
and their [OIII] $\lambda$5007/H$\beta$ flux ratio is lower than is
typical of normal Seyfert 1s (Shuder \& Osterbrock 1981; Osterbrock \&
Pogge 1985). The conventional definition for a NLS1 is a Seyfert 1
with H$\beta$ FWHM$<$2000 km s$^{-1}$ and [OIII]
$\lambda$5007/H$\beta$ $<$3 (Goodrich 1989). Previous studies focused
on such kind of Seyfert galaxies and showed many intriguing
characteristics. For example, NLS1s were often found to have low
black hole masses (e.g. Wang \& Lu 2001; Bian \& Zhao 2004; Zhou et
al. 2006; Komossa \& Xu 2007; Zhu, Zhang \& Tang 2009, hereafter:
Zhu09). {\bf These may be systematically lower than predicted by the
M-$\sigma_{*}$ relation which holds well for BLS1s (Grupe \& Mathur
2004; Mathur \& Grupe 2005).}  The NLS1s also have high Eddington
ratios (Boroson 2002; Komossa 2008; Paper-I).  It is thus proposed
that NLS1's central black hole may still be growing (e.g. Mathur,
Kuraszkiewicz, \& Czerny 2001; Komossa \& Mathur 2001; Komossa 2008).
In addition, NLS1s have softer 2-10 keV spectra, lower
2-10 keV luminosities, higher $\alpha_{ox}$ values and more energetic
BBB (see Paper-I and references therein).  In this paper, we continue
to pay special attention to the NLS1 subset in our sample and show
how they behave differently from other sources in the
cross-correlation study.

This paper is organized as follows. We first review some most important
characteristics of the sample in Section 2, in order to emphasize 
that our study conducted in the following sections should be 
related to the most intrinsic properties of AGN's bare core.
Section 3 will present the `Optical to X-ray Correlation Spectrum (OXCS)'
based on our new `Correlation Spectrum Technique (CST)', from which various
correlation features related to the hard X-ray luminosity are found 
in the optical spectrum.
Section 4 will study the cross-correlation between different 
Balmer line components and broadband SED components.
Section 5 will focus on correlations related to the Balmer line equivalent width 
(EW). Section 6 will study the physical properties of 
different Balmer emission line regions. [OIII] $\lambda$5007 line's 
property and its correlation with different SED components are put in
Section 7. Summary and conclusion will be made in Section 8. Following
Paper-I, flat universe model is adopted with the Hubble constant
H$_{0} = 72$ km s$^{-1}$ Mpc$^{-1}$, $\Omega_{M} = 0.27$ and
$\Omega_{\Lambda} = 0.73$.

\begin{figure*}
\centering
  \begin{tabular}{cc}
   \includegraphics[scale=0.48,clip=]{sed_example_PG1115p407.ps}&\includegraphics[scale=0.35,clip=1]{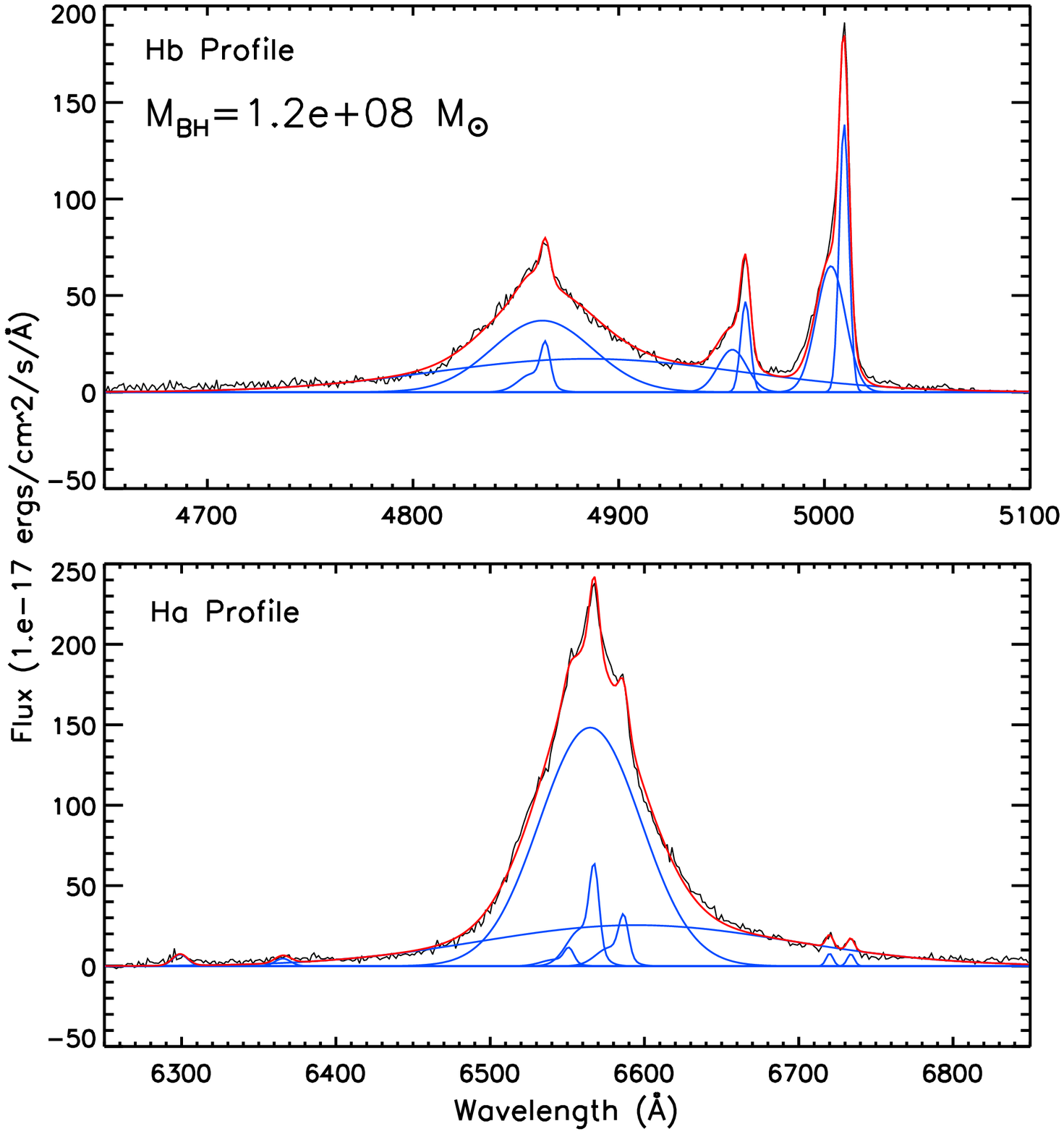}\\
  \end{tabular}
  \caption{Examples of spectral fitting in Paper-I. Left panel shows the broadband SED fitting of PG 1115+407 which consists of a modified accretion disc (green dashed line), a soft X-ray Comptonisation (orange dotted line) for the soft X-ray excess and a hard X-ray Comptonisation (blue dash-dotted line) for the hard X-ray power law tail.
  Right panel shows the emission line fittings of RBS 1423 around H$\alpha$ and H$\beta$. Blue solid lines represent different line components.}
  \label{fig:sed_example}
\end{figure*}

\section{The Sample and The Spectral Modelling}
\label{section:the sample}
\subsection{Sample Selection}
The sample used in this paper is a nearby unobscured Type 1 AGN sample
derived from the cross-correlation of 2XMMi \& SDSS DR7 catalogs.
{\bf The main selection steps are listed below for completeness. A
full source list
and more detailed explanation of the sample selection can be found in
Paper-I.\\ (1) We searched the 2XMMi and SDSS DR7 catalogs and identified
3342 extragalactic sources having both X-ray and optical spectra.\\ (2)
Within these sources, we selected those with $H\beta$ in emission and
redshift $z < 0.4$, so that both the H$\alpha$ and H$\beta$ emission
lines are covered by the SDSS spectra. This assists with modelling
of the Balmer lines (see Paper-I).  This selection resulted in 802
unique X-ray sources.\\ (3) Within this sample set, we identified 96
Type 1 AGNs all with a minimum of 2000 counts in at least one of the three
XMM-Newton EPIC cameras, to ensure high X-ray spectral quality. \\ (4)
We then excluded 23 sources whose $H\beta$ line was modified due to
high reddening, low S/N or a data gap in the SDSS spectra. The
remaining sample contains 73 AGNs.\\ (5) For each of the 73 sources, a
power law model was fitted to the rest-frame 2-10 keV X-ray
spectra. The 16 objects with photon index uncertainties greater than 0.5
were thereby excluded, leaving 57 Type 1 AGNs with relatively well
constrained 2-10 keV spectra.\\ (6) A further 6 objects were excluded
because of the obvious signature of a warm absorber at $\sim$0.7 keV.\\
The final sample contains 51 AGNs, with 12 AGNs being classified as
NLS1 using the conventional definition (Goodrich 1989), while the
others are all BLS1s.}  Most objects in this sample are radio quiet,
except for 3 sources that were reported as radio loud, i.e. PG
1004+130, RBS 0875 and PG 1512+370.
High quality XMM-Newton EPIC X-ray spectra and SDSS optical spectra are
available for every source in this sample. In addition, simultaneous
optical/UV photometric data from the XMM-Newton OM monitor are
available for 37 sources.

{\bf We exclude PG 1004+130 from all correlations as it is a BAL quasar, so
its X-ray flux is likely to be heavily obscured even though it does
not show clear evidence for absorption edges (Miller et al. 2006).
We also exclude Mrk 110
from the optical correlations (Sections 3, 4 and 5) as this source shows
strong optical variability (Kollatschny et al. 2001; Kollatschny 2003)
and the SDSS spectrum is very different from the (non-simultaneous)
XMM-Newton OM data (see Paper-I).
However, the [OIII] line luminosity does not
change with the optical continuum, so we include this object in the
[OIII] versus broad band SED correlations in Section 7.}

\subsection{Selection Bias}

{\bf Our sample distributes evenly within $0.031<z<0.377$ with a mean
redshift $<z>=0.137^{+0.158}_{-0.073}$.  Its selection is mainly based
on both high quality optical and X-ray spectra, so any AGN that was
not detected, or only marginal detected, by SDSS or XMM-Newton would not
be included in the sample, i.e. sources with low mass accretion rate
or strong obscuration were excluded.} Further selection criteria are
more related to specific spectral characteristics, such as excluding
objects with heavy optical reddening, Type 2 objects and X-ray warm
absorber objects.  Therefore it will be hard to directly estimate the
bias due to these selection effects. However, we could compare our
sample's general properties with previous samples. We found that our
sample had very similar redshift, 2-10 keV luminosity and bolometric
luminosity distributions as Vasudevan \& Fabian (2007)'s sample
(hereafter: VF07), except that VF07's sample also includes some
extremely nearby and low X-ray luminosity sources such as NGC 4395,
NGC 3227 and NGC 6814, which did not fall into our sample. {\bf The
redshift distribution of our sample is similar to that of Grupe et
al. (2010), except that their sample has a higher fraction of lower
redshift sources ($<z>=0.112\pm0.077$).}  The selection effect
regarding the broadband SED shape should be weak, which is because the
broadband SEDs of our sample have shown a very strong shape diversity,
with the intrinsic optical to X-ray spectral index $\alpha_{ox}$
ranges 1$\sim$2.  It is true that any objects with extraordinary
optical to X-ray luminosity ratios would not be included in our
sample. But such odd broadband SEDs are more likely due to optical or
X-ray obscuration, rather than being intrinsic to AGN, thus they are
no longer the type of unobscured sources we need in this sample.

\subsection{Major Sample Properties}

The most important characteristic of this particular sample is the
high quality of both their optical and X-ray spectra. In the optical,
none of these sources suffer strong dust reddening, thus all sources
have very clear optical underlying continuum superposed by a series of
clear broad and narrow emission lines. In the X-ray our selection
criteria have excluded sources whose spectra from XMM-Newton have low
signal to noise or contain strong warm absorber features (i.e. the
absorption edge at $\sim$0.7 keV, from combined absorption features
from partially ionized Oxygen and Iron, see e.g. Lee et al. 2001;
Turner et al. 2004).  The rest of the sources all have high quality
X-ray spectra which represent the emission from the AGN's core
emission.

\subsection{The Spectral Modelling}
\label{section:spectral modelling}
Another important characteristic is the availability of all important
spectral parameters from optical to X-ray for the whole sample, which
results from our thorough modelling of the multi-component Balmer
lines, optical spectrum and broadband SED.

In the emission line fitting (e.g. Figure~\ref{fig:sed_example} right
panel), two Gaussian components were used to fit the [OIII]
$\lambda$5007 line, i.e. a central component and a blue
component. {\bf Then the whole profile of [OIII] $\lambda$5007,
i.e. including both central and blue components, was used in the
fitting of the narrow component of Balmer lines.}  Two additional
Gaussian profiles were included in fitting each of the H$\alpha$ and
H$\beta$ lines, so that each line contains a narrow component (NC), an
intermediate component (IC) and a broad component (BC).  All other
strong nearby emission lines, e.g. [NII] $\lambda\lambda$6585/6548
doublets, Li $\lambda$6708, [SII] $\lambda\lambda$6717/6734, are
included by adding more Gaussian profiles into the whole model.
Various constrains were set for these Gaussian components which are
all described in Paper-I.

In the broadband SED fitting (e.g. Figure~\ref{fig:sed_example} left
panel), we made use of a new SED model ({\it optxagn\/} model in {\tt
Xspec} v12: Done et al. 2011), which modifies the accretion disc
emission (the green dashed line) by assuming a corona radius within
which all accretion disc emission is transferred to a soft X-ray
Comptonisation component (the orange dotted line) to account for the
observed soft X-ray excess, plus a hard X-ray Comptonisation component
(the blue dash-dotted line) to model the hard X-ray power law tail.
We rebuilt the broadband SED from the optical to hard X-ray by
extrapolating the best-fit model over the unobservable UV/soft X-ray
region, and then derived all the SED parameters. Detailed descriptions
of these spectral fitting can be found in Paper-I.

\section{The Optical to X-ray Correlation Spectrum (OXCS)}
\label{section:OXCS}

\subsection{The Motivation of OXCS}
The hard X-ray emission from AGN rises from Compton up-scattering of
disc photons by a high temperature (100s of keV) electron population
which forms a corona region located above the accretion disc.
Although this hard X-ray emission may only contribute a small fraction
of the total central ionizing flux (depending on the corona radius,
see Paper-I), it is capable of penetrating deeply into the most dense
gas regions near AGN's core, which emits some specific line
species. Provided that the intrinsic hard X-ray emission is not
heavily absorbed and the optical spectrum is not heavily reddened, we
can identify those optical emission features closely linked with the
high energy core emission by testing their correlation with the hard
X-ray.  As noted above, our sample was selected based on the
unobscured nature of both the optical and X-ray spectra, so the hard
X-ray luminosity can be used as a reliable diagnostic to investigate
its relation to the optical spectral properties. Here we propose and
test a new type of spectrum, the `Optical to X-ray Correlation
Spectrum (OXCS)' (see Figure~\ref{fig:oxcs}).  {\bf This is a direct
extension of previous monochromatic luminosity correlation studies
between the X-ray and optical e.g. L$_{2keV}$ vs. L$_{2500\AA}$
(e.g. Green et al. 2009; Lusso et al. 2010).}

\subsection{Construction of OXCS}
\label{section:OXCS construction}
\begin{figure*}
\centering
\includegraphics[scale=0.7,angle=90,clip=]{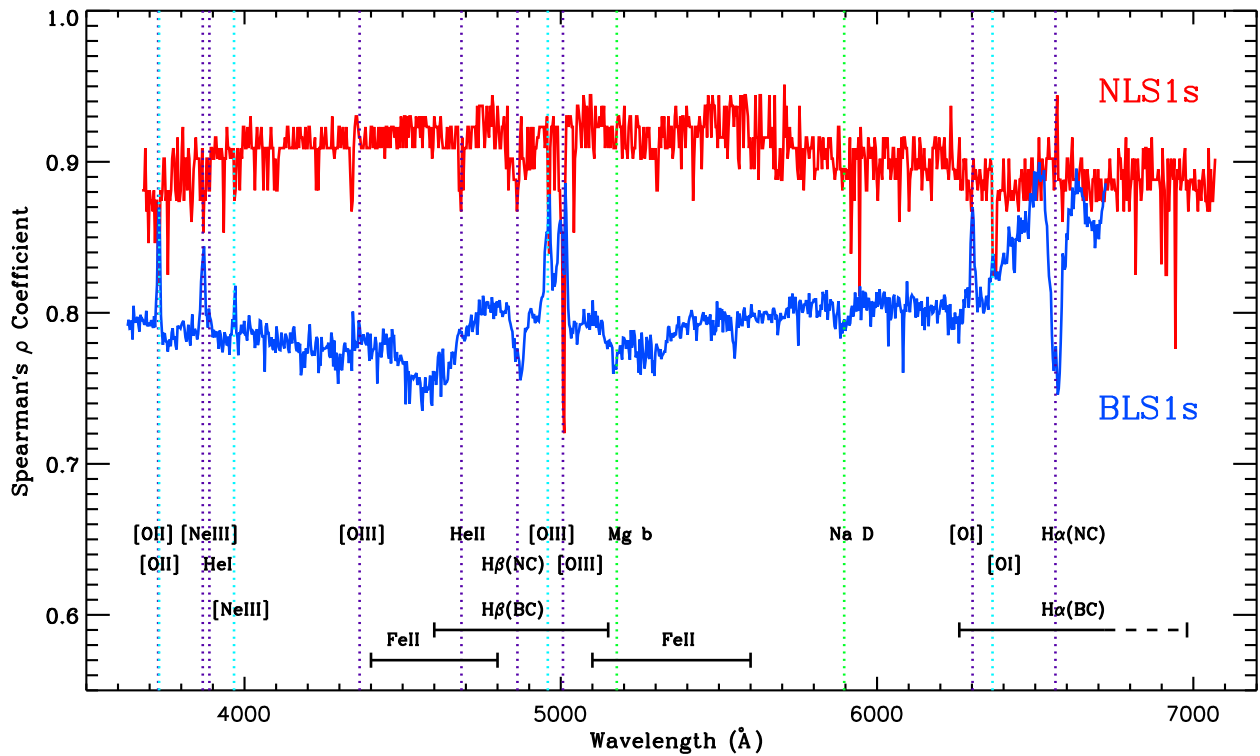}
\caption{{\bf The OXCSs for our sample, the method of constructing them is described
in Section~\ref{section:OXCS construction}.
The red line is the OXCS for the 12 NLS1s in our sample. The blue line is the OXCS for the
37 BLS1s in our sample. Purple and cyan dotted lines indicate
the wavelengths of some most prominent optical emission lines for a typical AGN,
with cyan lines indicating the weaker line of any doublets.
Green dotted lines indicates the wavelengths of Mg {\it b\/} and Na {\it D\/}
stellar absorption features. `(NC)' is the narrow component, while `(BC)' is the broad
component. The dashed region of H$\alpha$(BC) means that this region is not covered
by the BLS1 OXCS.}}
\label{fig:oxcs}
\end{figure*}

The principle behind the OXCS is to cross-correlate the hard X-ray
luminosity (here we choose the luminosity of 2-10 keV: L$_{2-10keV}$)
with the monochromatic luminosities at each wavelength of the optical
spectrum for the whole sample of objects. Then we plot the correlation
coefficient against the wavelength, to see how the correlation changes
with wavelength.  For each source in the sample, we corrected the SDSS
spectra for Galactic reddening and de-redshifted them to their rest
frame. We define a standard optical spectral region that is covered by
the SDSS spectrum of every source (around 3700-6700\AA), and calculate
the monochromatic luminosity at 1000 wavelengths distributed evenly
across this spectral range.  The Spearman's rank test was used to
cross-correlate these monochromatic luminosities with L$_{2-10keV}$,
and so the Spearman's $\rho$ coefficient was derived at each of the
1000 wavelengths.  Figure~\ref{fig:oxcs} plots the Spearman's $\rho$
coefficient against the wavelength for the 12 NLS1 (red line) and 
37 BLS1 (blue line). 
The wavelengths of some of the most prominent
optical emission and absorption features in a typical AGN spectrum are
indicated in the plot.  Note that the spectral coverage is not exactly
the same for different OXCS subsets, because these subsets
have slightly different redshift ranges.

\subsection{Correlation Features}
\label{section:oxcs:correlation features}
Within the limited wavelength range of OXCS, the underlying continuum
correlation would not be expected to change significantly, and so it
forms a basic correlation level that is relatively flat in OXCS.
Superposed on this basic correlation continuum there are various
emission and absorption-line-like features, which shows that those
lines have stronger or weaker correlation with L$_{2-10keV}$ than the
underlying continuum. We identify some of the most noticeable
characteristics in the OXCSs as below:\\ 
\\ 
(1) L$_{2-10keV}$ emission
correlates well with the entire optical underlying continuum.  {\bf
The underlying correlation is $\sim$0.8 in the BLS1 OXCS, and
$\sim$0.9 in the NLS1 OXCS.} A noticeable phenomena is that the
underlying correlation does not decrease significantly towards either
the blue or red end of the optical spectral range.  This confirms that
our sample suffers little intrinsic reddening, and that the
optical continua redward of 5000{\AA} are dominated by AGN emission,
i.e. the host galaxy contamination is small in most cases.
There is also suggestions about the existence of an extra component
contributing optical emission at redward of 5000\AA, which is
probably originated from the self-gravity dominated region
of accretion disc (Vanden Berk et al. 2001; Collin \& Hur\'{e}
2001; Puchnarewicz et al. 2001; Soria \& Puchnarewicz 2002;
Pierens, Hur\'{e} \& Kawagushi 2003; Collin \& Kawaguchi 2004;
Hao et al. 2010; but see Landt et al. 2011). But we find it
difficult to investigate this component merely using SDSS
spectra due to the difficulties in accurately subtracting the
host galaxy emission.
\\ 
(2) {\bf In the
BLS1 OXCS, the broad wings of H$\alpha$ and H$\beta$ correlate better
with L$_{2-10keV}$ than the optical continuum,} and so result in the
apparent broad-wing-like features around 4860{\AA} and 6565{\AA}.
However, the core region of Balmer lines has a much weaker correlation
with L$_{2-10keV}$ as shown by the two narrow correlation dips centred
at 4862.68{\AA} and 6564.61{\AA} in the OXCSs. This directly shows
that the Balmer lines in BLS1s consist of (at least) two components, from the
NLR and BLR of different physical conditions.  We will
investigate this issue further in later sections.  The [OIII]
$\lambda\lambda$4959/5007 doublets in BLS1s show a very strong
correlation with L$_{2-10keV}$, in spite that the [OIII]
$\lambda\lambda$4959/5007 originates in the NLR far from AGN's core. We
will investigate this in more detail in
Section~\ref{section:OIII}. The Balmer line profile in NLS1s is not
strongly broadened, {\bf so the NLS1 OXCS does not exhibit similar
correlation bumps around Balmer lines as seen in the BLS1 OXCS.}  In
contrast to the BLS1s, the [OIII] $\lambda\lambda$4959/5007 in NLS1s
also have much weaker correlation with L$_{2-10keV}$ than their local
optical continuum.\\ 
\\ 
(3) {\bf The BLS1 OXCS also exhibit
emission-line-like features at} the wavelengths of some other emission
lines in a typical BLS1 optical spectrum, such as [NeIII]
$\lambda\lambda$3869/3967, [OI] $\lambda\lambda$6300/6364 and [OII]
$\lambda\lambda$3726/3729. This suggests that these emission lines all
correlate strongly with the hard X-rays. However, [OI] $\lambda$6300 is a
relatively weak line, and its prominence in the OXCS may support the
existence of dense gas clouds near AGN's core inside which gas stays
neutral or at low ionization.  Only hard X-rays can penetrate into these
clouds and produce such low ionization lines.  The fact that [OII]
$\lambda\lambda$3726/3729 correlates quite well with L$_{2-10keV}$
suggests that reddening is indeed quite low for our sample, since
otherwise the presence of dust would tend to diminish any correlation.
It is apparent that in terms of the OXCS around emission lines, the
NLS1s are different from BLS1s.  This may be a result of geometrical
effects, or that the line emitting regions in NLS1s are not as closely
associated with hard X-ray emission as in the BLS1s.\\ 
\\ 
(4) On the
contrary, it is seen that the stellar absorption lines Mg {\it b\/}
and Na {\it D\/} do not correlate well with L$_{2-10keV}$, producing
absorption-like features in the OXCSs.  Neither does FeII emission in
the ranges 4400-4800 {\AA} and 5100-5600 {\AA} correlate with
L$_{2-10keV}$, especially for BLS1s.

\subsection{The Correlation Spectrum Technique (CST)}

More generally, the OXCS provides a new tool for spectral
studies based on medium to large samples of objects. We will name 
this the `Correlation Spectrum Technique (CST)'.
As shown above, an example of the CST is the OXCS,
which has proved to be useful for investigating hard X-ray related correlations
in the optical spectrum for different AGN populations.

Using hard X-ray as a diagnostic, we can also apply the CST to
spectra of longer wavelengths such as near and far infrared,
provided that an AGN sample with good spectral data is available.
We can also use luminosities other than hard X-ray
as the diagnostic in CST. For example, in Section~\ref{section:SOCS} we constructed
the `SED to [OIII] $\lambda$5007 Correlation Spectrum (SOCS)', in which case
the CST uses the luminosity of [OIII] $\lambda$5007 to produce the
correlation spectrum from optical to hard X-ray for different sample subsets.

\section{Balmer Line Luminosity}
\label{section:balmer:lum}

\begin{figure*}
\centering
\includegraphics[bb=0 300 596 842, scale=0.8, clip=1]{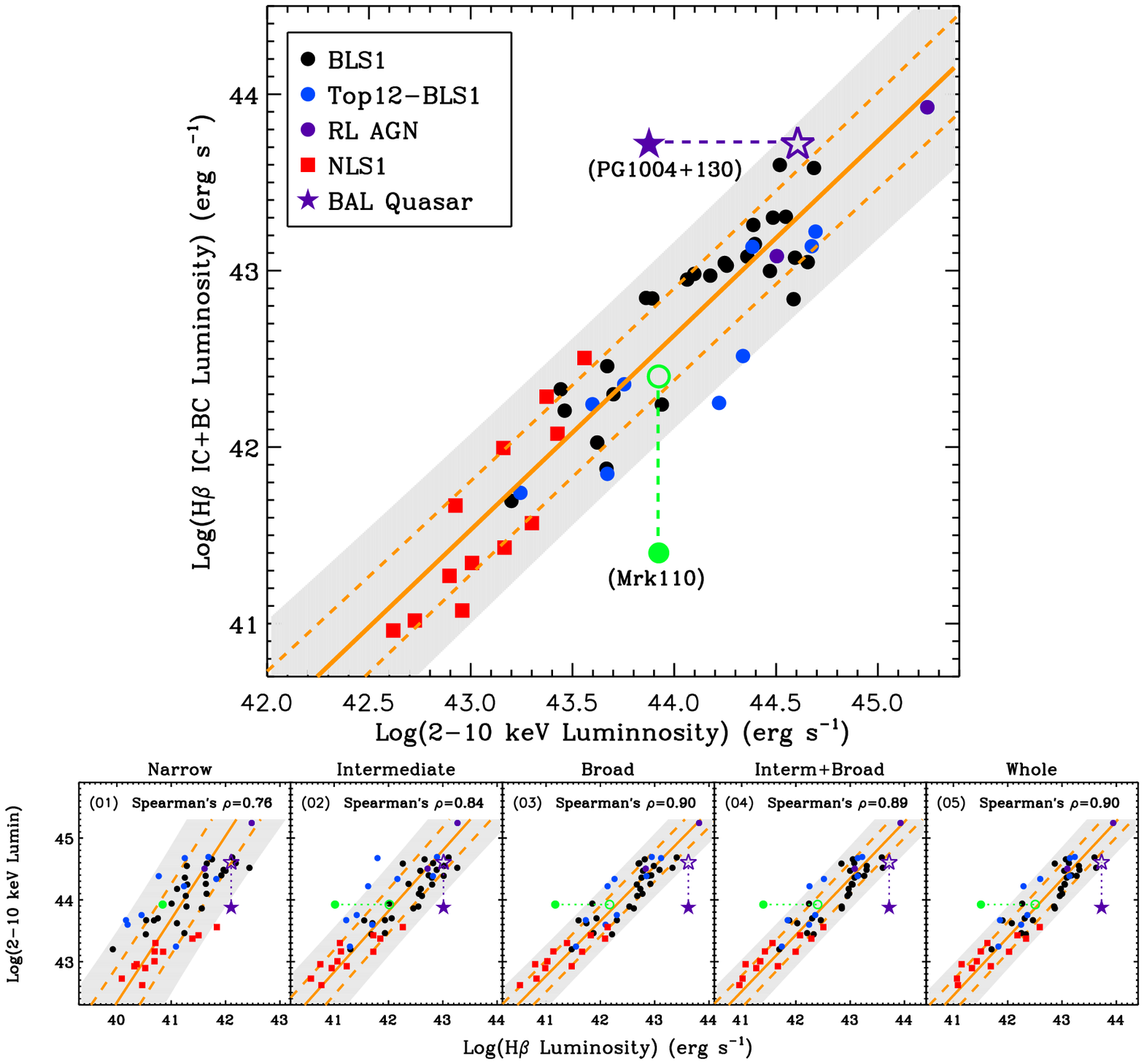}
\caption{The luminosity correlation between Balmer line components 
and 2-10 keV. The upper figure shows H$\beta$ luminosity (NC subtracted) 
vs. L$_{2-10keV}$. The connected filled and empty 
purple stars indicate the position of PG 1004+130 before and after being
corrected for the 0.73 dex (Miller et al. 2006). The connected filled
and empty green circles indicate different optical positions of Mrk 110 
as calculated from the SDSS spectrum and the FAST spectrum (Landt et al. 2011).
The solid orange line shows the linear regression line treating L$_{2-10keV}$
as the independent variable, with the two dashed orange lines indicating 
the $\pm$1$\sigma$ region for new observations, and the shaded region
showing the $\pm$2$\sigma$ region. The lower
panels present the same type of correlations for different H$\beta$ 
components, i.e. H$\beta$ NC, IC, BC, IC+BC (or NC-sub) and the whole
line. In each plot, Spearman's rank coefficients were calculated after
excluding PG 1004+130 and Mrk 110. The regression coefficients are listed 
in Table~\ref{app:table:spearman}.}
\label{hblum:210lum:plot}
\end{figure*}

For Type 1 AGN, each Balmer line profile consists of two to three distinct components, 
i.e. narrow component (NC) from the narrow line region (NLR) which extends a few 
hundred parsecs from the central black hole
(e.g. Bennert et al. 2006; Heckman et al. 2005), broad component (BC) from the
broad line region (BLR) which is tens of light-days from the black hole 
(e.g. Kaspi et al. 2005; Bentz et al. 2006). Sometimes  
another intermediate component (IC) is present, with moderate linewidth which is probably
originated from the intermediate line region (ILR) which may extend up to the region
of the inner radius of the dusty torus (e.g. Zhu09). Therefore the distance of these line 
emitting regions from the compact AGN goes as NLR, ILR, BLR from the farthest to
the closest. Among these emission line regions we may also expect a density
gradient and a correlation trend with the central ionizing flux.
The correlation between the Balmer line luminosity and X-ray
luminosity has been known long ago (e.g. Ward et al. 1988),
but the hard X-ray correlations for different Balmer line
components have never been studied.
Now our sample with both high quality optical and X-ray
spectra provides an opportunity for this study.

\begin{figure*}
\centering
\includegraphics[bb=340 72 540 792, scale=0.7, clip=1,angle=90]{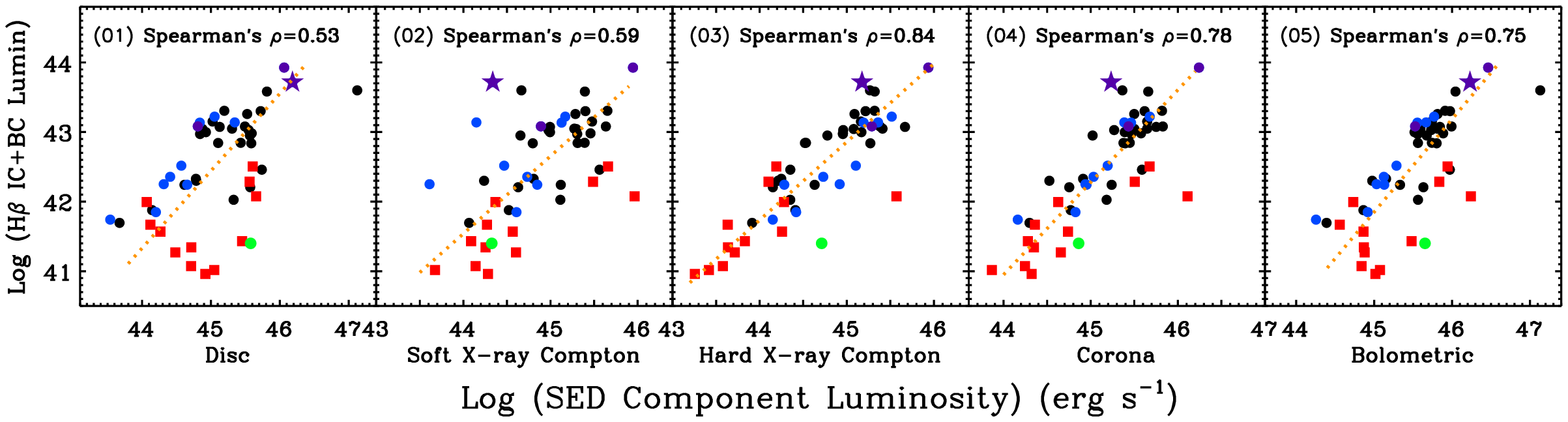}
\caption{The luminosity correlations between H$\beta$ IC+BC and broadband SED components.
`Corona' means the coronal luminosity,
which is the sum of the luminosities of the soft and hard X-ray Comptonisation components.
Different symbols represent different type of sources as explained in Figure~\ref{hblum:210lum:plot}.
In each panel the Spearman's rank coefficient is given, along with the
orange dotted line indicating the bisector regression line.}
\label{fig:hbbc-sed-lum}
\end{figure*}

\subsection{Balmer Line Component Luminosity vs. L$_{2-10keV}$}
\label{hblum:210lum:section}

We have shown in the previous section that different components in
the H$\alpha$ and H$\beta$ lines may have different correlation status
with L$_{2-10keV}$. In this section we investigate the correlation of
L$_{2-10keV}$ with the individual components (Narrow, Intermediate and
Broad) derived from our Balmer line decompositions for each source in
Paper-I. The best correlation is for L$_{2-10keV}$ vs. H$\beta$
IC+BC (i.e. the combination of NC and BC, equivalent to NC subtracted
H$\beta$) luminosity, shown in the upper panel in Figure~\ref{hblum:210lum:plot}.
NLS1 (red squares) and BLS1 (blue circles)
clearly lie on the same strong correlation. Spearman's rank test gives
a correlation coefficient of $\rho_{s}=0.9$ and probability of
random distribution of 
$d_{s}=8.1\times10^{-19}$. 

We also plot the uncorrected data from our two excluded sources (PG
1004+130: purple star and Mrk 110: green circle) on the
correlation. These strongly deviate from the best-fit line (orange
solid line), the $\pm$1$\sigma$ lines (orange dash lines) and the
$\pm$2$\sigma$ region (light gray region). {\bf However, PG 1004+130
is the only BAL quasar in our sample, whose X-ray emission after
correcting for intrinsic absorption, was reported as 0.73 dex weaker than
normal PG RLQs normalized to similar optical/UV luminosities
(Miller et al. 2006). Mrk 110's optical continuum is highly variable
(Kollatschny et al. 2001; Kollatschny 2003). The SDSS spectrum of
Mrk 110 is $\sim$1 order of magnitude less luminous than the {\sc FAST}
optical spectrum (Landt et al. 2011),} which is just the required
amount of correction we need to pull Mrk 110 back onto the best-fit
correlation line. Therefore, we conclude that this tight correlation
between H$\beta$ IC+BC luminosity and 2-10 keV luminosity is an
intrinsic property. We propose that if an AGN is found to strongly
deviate from this correlation, then it is likely that its X-ray or
optical emission is obscured.

We calculated three types of regression lines as described in Isobe et
al. (1990).  Since H$\alpha$ also gives similar results (see
Table~\ref{table:balmerlum:210lum}), we give bisector regression
equations for both H$\beta$ and H$\alpha$:\\ (i) L$_{2-10keV}$
expressed by the Balmer line Luminosities:
\begin{eqnarray}
\label{balmerlum:210lum:eqn:1}
Log L_{2-10}=(0.83{\pm}0.03)Log L_{H\alpha(NCsub)} + (8.35{\pm}1.43)\\
Log L_{2-10}=(0.83{\pm}0.04)Log L_{H\beta(NCsub)} + (8.56{\pm}1.52)&
\end{eqnarray}
(ii) Balmer line Luminosities expressed by L$_{2-10keV}$ :
\begin{eqnarray}
\label{balmerlum:210lum:eqn:2}
Log L_{H\alpha(NCsub)}=(1.20{\pm}0.05)Log L_{2-10} - (9.50{\pm}2.09)\\
Log L_{H\beta(NCsub)}=(1.18{\pm}0.04)Log L_{2-10} - (9.17{\pm}2.04)&
\end{eqnarray}
Using the above equations we estimate that PG 1004+130's X-ray luminosity
is weaker than normal Type 1 AGNs by 1.0$\pm$0.3 dex,
which is slightly higher than Miller et al. (2006)'s estimation
of 0.73 dex weaker than normal PG radio-loud quasars with similar optical/UV
luminosities based on their X-ray spectral analysis.

\begin{table}
 \centering
   \caption{The luminosity correlations between Balmer line components and 2-10 keV. The Bisector regression coefficients in the equation: Log($L_{BalmerLine}$)=$\beta_{0}$Log($L_{2-10keV}$)+$\xi_{0}$ are listed, along with the Spearman's rank correlation coefficients ($\rho_{s},~d_{s}$) as defined in Table~\ref{app:table:spearman}.}
    \begin{tabular}{@{}lcccc@{}}
    \hline
    Line Comp. & \multicolumn{2}{c}{Bisector Regress Coef.} & \multicolumn{2}{c}{Rank Cor.}\\
    H$\alpha$ & $\beta_{0}$&$\xi_{0}$&$\rho_{s}$&$d_{s}$\\
    \hline
    NC vs. L$_{2-10keV}$&1.00$\pm$0.06&-2.01$\pm$3.28&0.76&-10\\
    \hline
    IC vs. L$_{2-10keV}$&1.18$\pm$0.05&-9.00$\pm$2.26&0.92&-20\\
    \hline
    BC vs. L$_{2-10keV}$&1.23$\pm$0.05&-11.51$\pm$2.19&0.93&-22\\
    \hline
    \multicolumn{2}{@{}l}{IC+BC vs. L$_{2-10keV}$}&&&\\
    (i.e. NC-sub)&1.20$\pm$0.05&-9.50$\pm$2.09&0.94&-22\\
    \hline
    Whole vs. L$_{2-10keV}$&1.15$\pm$0.04&-7.66$\pm$1.95&0.93&-22\\
    \hline
    H$\beta$ & $\beta_{0}$&$\xi_{0}$&$\rho_{s}$&$d_{s}$\\
    \hline
    NC vs. L$_{2-10keV}$&1.02$\pm$0.07&-3.49$\pm$3.35&0.76&-10\\
    \hline
    IC vs. L$_{2-10keV}$&1.17$\pm$0.05&-9.41$\pm$2.54&0.86&-15\\
    \hline
    BC vs. L$_{2-10keV}$&1.21$\pm$0.05&-10.67$\pm$2.09&0.93&-21\\
    \hline
    \multicolumn{2}{@{}l}{IC+BC vs. L$_{2-10keV}$}&&&\\
    (i.e. NC-sub)&1.18$\pm$0.04&-9.17$\pm$2.04&0.92&-20\\
    \hline
    Whole vs. L$_{2-10keV}$&1.15$\pm$0.04&-7.91$\pm$1.97&0.93&-21\\
    \hline
    \end{tabular}
    \label{table:balmerlum:210lum}
\end{table}

The lower panels in Figure~\ref{hblum:210lum:plot} show the (weaker
but still very significant) correlations for the different line
components. It is clear that the BC has the best correlation with hard
X-ray emission as found previously; IC+BC and the whole H$\beta$ line
also show good correlations; the NC related correlations are not as
good as others.  This confirms that the BLR has the closest link with
the AGN's central X-ray continuum.

We note that the well-known Malmquist bias (Gonzalez \& Faber 1997,
and references therein) will be partly responsible for these
correlations, but it should affect all of the correlations equally.
Hence the change in correlation strengths among the different Balmer
line components should be real. We also examined these correlations
using flux instead of luminosity, and a very similar trend of strength
of correlation was found for NC, IC and BC.  Note that the components
in H$\alpha$ show similar correlation status as in H$\beta$ (see
Table~\ref{table:balmerlum:210lum}).

\subsection{Cross-correlation between Balmer Line Components and Broadband SED Components}

In Paper-I we decomposed the Balmer lines into broad, intermediate and
narrow components. We also decomposed the broadband SED into three
components, namely the disk, soft X-ray Comptonisation and hard X-ray Comptonisation.
Thus it is possible for us to correlate each line component with each SED component.
The results are presented in Table~\ref{app:table:spearman} and
Figure~\ref{app:hblum:sedlum:plot}. {\bf Note that for each source
we subtracted the FeII emission from the nearby region of the H$\beta$
line before conducting the profile decomposition, so the dispersion in
these correlations is a result of FeII contamination.}  We find
that among the three SED components, the hard X-ray Comptonisation
produces the best correlations, while the accretion disc emission and
soft X-ray Comptonisation show weaker correlations.  Among the
different Balmer line components, the correlation strengthens from NC,
IC to BC.  Figure~\ref{fig:hbbc-sed-lum} shows the correlation status
for H$\beta$ IC+BC vs. broadband SED components. It is clear that the
best correlation is found in the hard X-ray component though the
correlation is also good for adding both soft and hard X-ray
Comptonisation together as total corona luminosity.

Figure~\ref{fig:hbbc-sed-lum} also shows that for BLS1s the H$\beta$
IC+BC correlate well with the accretion disc luminosity and bolometric
luminosity.  However, NLS1s are much more dispersed in these
correlation plots, and thus dilute the correlation strength of the
whole sample. We need a larger sample to confirm the different
behaviors of NLS1s and BLS1s in these correlations.

\begin{figure*}
\centering
\includegraphics[scale=0.7, clip=1,angle=90]{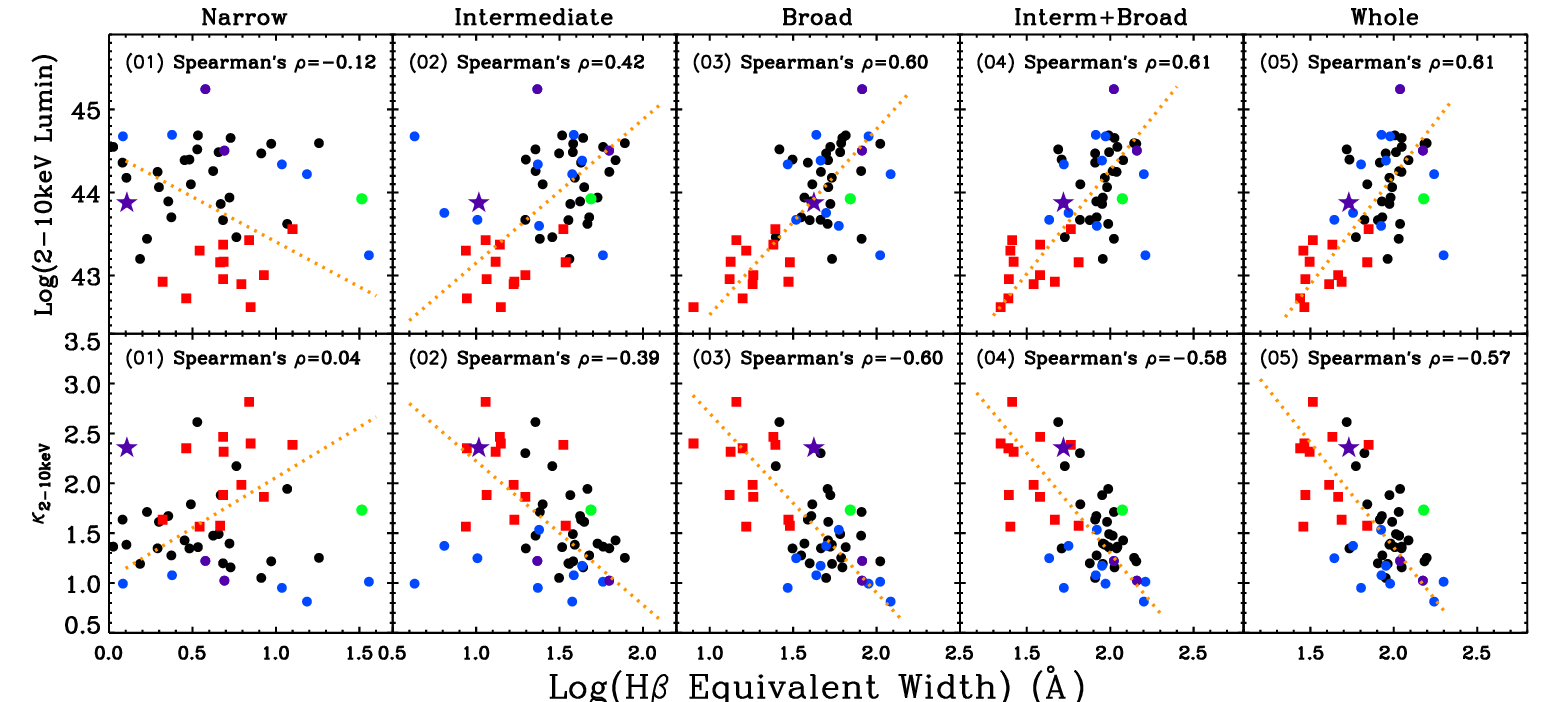}
\caption{The correlations of H$\beta$ component EW vs. L$_{2-10keV}$ 
(first row), $\kappa_{2-10keV}$ (second row). 
Different symbols represent the same type of sources as in
Figure~\ref{hblum:210lum:plot}. Spearman's rank coefficients are
calculated for the whole sample.
The orange dotted line indicates the bisector regression line.}
\label{hbew:210kev:plot}
\end{figure*}

\section{Balmer Line Equivalent Width (EW)}
\label{section:balmer:EW}

\subsection{Balmer Line Component EW vs. L$_{2-10keV}$}
In the previous section we reported the strong correlation between
H$\beta$ luminosity and L$_{2-10keV}$ as
L$_{H\beta}~{\propto}~$L$_{2-10kV}^{1.15\pm0.04}$ for the whole
H$\beta$ line, and
L$_{H\beta(IC+BC)}~{\propto}~$L$_{2-10kV}^{1.18\pm0.04}$ for the IC
plus BC in H$\beta$. The index 1.18$\pm$0.04 is bigger than unity
which indicates that the luminosity of broader components of H$\beta$
increase faster than linearly with L$_{2-10kV}$ (also see
Table~\ref{table:balmerlum:210lum}). The situation is the same for H$\alpha$.
In order to investigate this
issue further, we study the properties of Balmer line EW.  We perform
the cross-correlation between L$_{2-10keV}$ and H$\beta$ component EW.
Figure~\ref{hbew:210kev:plot} shows the results.  There is no
correlation between L$_{2-10keV}$ and H$\beta$ NC EW, but the
correlations between L$_{2-10keV}$ and H$\beta$ IC, BC EWs are
significant as confirmed by Spearman's rank test (see
Table~\ref{app:table:spearman}). A bisector regression analysis shows
that H${\beta}$ IC+BC EW ${\propto}~$L$_{2-10kV}^{0.45\pm0.03}$.

{\bf We also find clear anti-correlations between the 2-10 keV
bolometric correction
(i.e. $\kappa_{2-10keV}$=L$_{bol}$/L$_{2-10keV}$) and the H$\beta$ EWs, as
shown in the second row of Figure~\ref{hbew:210kev:plot}.}
$\kappa_{2-10keV}^{-1}$ is the fraction of 2-10 keV emission in the
Bolometric luminosity, thus these correlations suggest that as the
fraction of 2-10 keV emission increases, the EWs of H$\beta$ IC and BC
also increase. This is not surprising since there is no correlation
between H$\beta$ EWs and the bolometric luminosity, as shown in
Table~\ref{app:table:spearman} the Spearman's rank correlation
coefficients ($\rho_{s}$) are only -0.08, 0.13, 0.21, 0.22 and 0.22
for L$_{bol}$ vs. the EW of H$\beta$ NC, IC, BC, IC+BC and the whole
line, respectively. The sources with large $\kappa_{2-10keV}$ are mostly NLS1s
which tend to have high mass accretion rates in terms of Eddington,
L$_{bol}$/L$_{Edd}$ (Vasudevan \& Fabian 2007, 2009; Paper-I; Jin et
al. in prep., hereafter: Paper-III).  However, when we directly
cross-correlated Eddington ratio with the various H$\beta$ EWs, no
significant correlations were found (Figure~\ref{app:fig:balmerew}).
Similar results were found when using H$\alpha$ line instead of
H$\beta$.

\subsection{Does A Balmer Line Baldwin Effect Exist?}
\label{hb:baldwin:effect}
The existence of the H$\beta$ Baldwin effect is controversial.
It was reported by Zhu09 that the H$\beta$ IC EW anti-correlates with 
the monochromatic luminosity at 5100{\AA} (hereafter: L$_{5100}$), with Pearson rank 
correlation coefficient being -0.48 and at 99\% level of smaller than 0, 
while the BC EW and the whole NC subtracted H$\beta$ EW does not 
correlate with L$_{5100}$. They suggest this should be due to the flat 
geometry of ILR and spherical geometry of BLR, but they did not mention 
whether such anti-correlation could also be found in their H$\alpha$ IC or not.
To compare with Zhu09's results, we use our Balmer line fitting 
results to perform similar correlation test.

\begin{figure}
\centering
\includegraphics[bb=30 144 594 660, scale=0.45,clip=1]{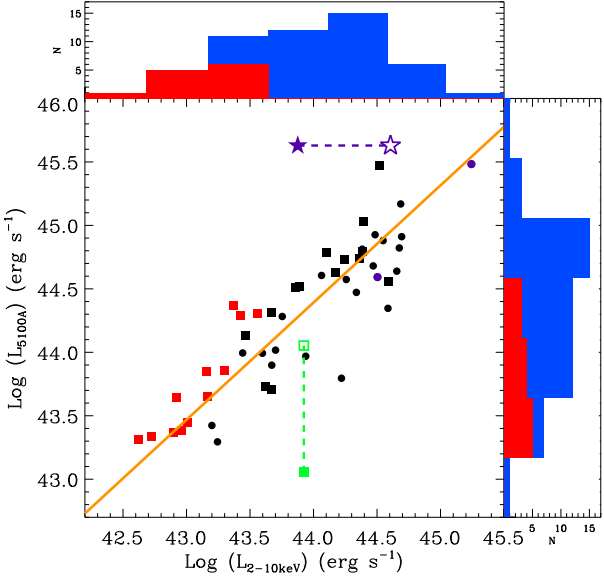}
\caption{L$_{2-10keV}$ vs L$_{5100}$. Different symbols represent the same type of sources as in Figure~\ref{hblum:210lum:plot}. Solid orange line is the bisector regression line assuming L$_{2-10keV}$ is the independent variable. In each histogram, the red region highlights the distribution of the 12 NLS1s in our sample.}
\label{fig:l210:l5100}
\end{figure}

We first investigate the correlation between L$_{2-10keV}$ and L$_{5100}$
which can be seen directly from the OXCSs in Figure~\ref{fig:oxcs}.
The L$_{2-10keV}$ vs. L$_{5100}$ correlation is plotted in Figure~\ref{fig:l210:l5100}
with Spearman's $\rho_{s}=0.88$ and $d_{s}=1.1\times10^{-16}$.
The bisector regression lines are found to be:\\
(i) L$_{5100}$ expressed by L$_{2-10keV}$:
\begin{equation}
\label{equ:l210:l5100:1}
Log(L_{5100})~=~(0.92\pm0.05)Log(L_{2-10keV})~+~(3.76\pm2.24)
\end{equation}
(ii) L$_{2-10keV}$ expressed by L$_{5100}$:
\begin{equation}
\label{equ:l210:l5100:2}
Log(L_{2-10keV})~=~(1.08\pm0.06)Log(L_{5100})~-~(4.07\pm2.66)
\end{equation}
Considering the strong correlations reported in previous paragraphs
between H$\beta$ IC, BC EWs and L$_{2-10keV}$,
the correlations between H$\beta$ IC, BC EW and L$_{5100}$ were expected.
However, we do not confirm any strong positive or negative correlations
between H$\beta$ IC, BC EWs and L$_{5100}$, though there is 
a large scatter in these cross-correlation plots,
as shown in Figure~\ref{app:fig:balmerew} and Table~\ref{app:table:spearman}.
There is a weak anti-correlation between NC EW and L$_{5100}$.
We therefore conclude that no evidence of Baldwin effect is found in our study
for Balmer IC and BC, but there may be such an effect for NC.

In fact, there are some important uncertainties in Balmer line
decomposition and L$_{5100}$, which need to be considered before
studying the Baldwin Effect of Balmer line components:\\ 
(1) The spectral quality of the Balmer line profile is crucial since currently
decomposition of Balmer line totally depends on the line profile.
This is more of a problem for H$\beta$ since strong reddening can
significantly reduce its S/N and distort its profile, so the H$\beta$
decomposition for such reddened sources will be unreliable. But this
is not a problem for our low reddening sample.\\ 
(2) Even with high
quality Balmer line profiles, the line decomposition still has
uncertainties.  It is highly probable that the Balmer lines in Type 1
AGN must contain at least a NC and a BC. Assuming a Gaussian or
Lorentzian profile for the BC, an additional IC is required during the
line fitting procedure by the $\chi^{2}$ statistics. However, the
assumption of Gaussian or Lorentzian profile is not secure. In some
cases, Balmer lines also exhibit a double-peak profile (Eracleous \&
Halpern 2003; Strateva et al. 2006; Bian et al. 2007) or an extended
flat red wing (e.g. Mrk 0926, see Paper-I), which causes problem for
the three-component decomposition.\\ 
(3) The L$_{5100}$ may not have
just one contributor.  In addition to the standard accretion disc
emission in the optical, L$_{5100}$ may also contain stellar emission
from host galaxy. It was reported that an additional component,
probably from the outer region of accretion disc where self-gravity
dominates, might also contribute a significant fraction of L$_{5100}$
(see Section~\ref{section:oxcs:correlation features}).\\ 
Therefore, it is difficult to find the intrinsic correlations between Balmer
line component EWs and L$_{5100}$ due to the above uncertainties, and so
the existence of Baldwin effect in any Balmer line component is still
unclear.

\subsection{Balmer Line Component EW vs. FWHM and BH Mass}
\label{hbew:fwhm:bhmass}
\begin{figure*}
\centering
\includegraphics[scale=0.7, clip=1,angle=90]{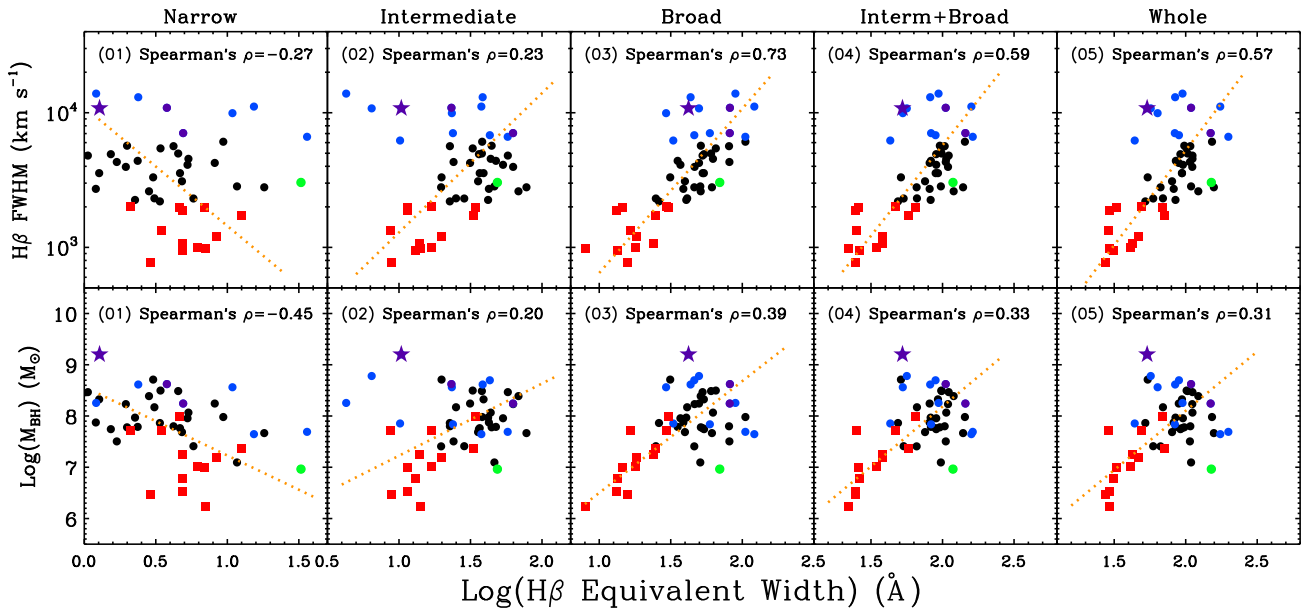}
\caption{The correlations of H$\beta$ component EW vs. H$\beta$ IC+BC 
FWHM (first row) and best-fit black hole mass (second row). Different 
symbols represent the same type of sources as in 
Figure~\ref{hblum:210lum:plot}. Spearman's rank coefficients were 
calculated for the whole sample.
The orange dotted line indicates the bisector regression line.}
\label{hbew:fwhm:bhmass:plot}
\end{figure*}

Another result we found for Balmer line EW is its correlation with Balmer line
width and black hole mass. We cross-correlate the H$\beta$ component EWs 
with the H$\beta$ FWHM$_{IC+BC}$ and the `best-fit' black hole mass
(see the description of `best-fit' black hole mass in Paper-I).
Figure~\ref{hbew:fwhm:bhmass:plot} shows our results. It is clear
that there are significant correlations between FWHM, black hole mass and 
the EWs of IC and BC. The best correlation is again found in BC.
The results suggest that as the black hole mass
increases (so does the Balmer line width), the emission from ILR and BLR becomes 
more luminous relative to the continuum luminosity. We also note from 
Figure~\ref{hbew:fwhm:bhmass:plot} that if only consider BLS1s (circular 
points in Figure~\ref{hbew:fwhm:bhmass:plot}),
then these is almost no correlation either between H$\beta$ FWHM and component EWs
or between black hole mass and H$\beta$ component EWs.
The broadest 12 BLS1s (blue circular symbols) exhibit 
larger scatter than the rest of the sources. 
However,  the correlation between H$\beta$ 
BC EW and black hole mass is very strong for NLS1s (red square symbols). 
There also seems to be a weak anti-correlation between black hole mass and 
H$\beta$ NC EW. Similar results can be found if replacing H$\beta$ with H$\alpha$.

\subsection{The Nonlinear Dependence of Balmer IC and BC Luminosities on L$_{2-10keV}$ and L$_{5100}$}

As shown in previous sections that the relations between
L$_{H\beta(IC+BC)}$, L$_{2-10keV}$ and L$_{5100}$ can be expressed as:
\begin{equation}
\label{equ:exp-dep}
L_{H\beta(IC+BC)}~{\propto}~L_{2-10keV}^{1.18\pm0.04}~{\propto}~L_{5100}^{1.28\pm0.05}
\end{equation}
Similar results can be found for H$\alpha$. Such non-linear dependences
imply that if the continuum luminosity is the first-order factor,
then there must be a second-order factor causing the EW of Balmer line IC and BC
to depend on L$_{2-10keV}$. This second-order factor
could be the covering factor of the ILR and BLR seen from the central ionizing continuum.
AGNs with higher L$_{2-10keV}$ and L$_{5100}$ may also have larger
ILR and BLR covering factors, making their Balmer IC and BC EWs larger 
than in the AGNs of low L$_{2-10keV}$ and L$_{5100}$.

\section{The Properties of ILR And BLR}
\subsection{Balmer Decrement}
\begin{figure}
\centering
\includegraphics[bb=90 360 414 920, scale=0.75, clip=1]{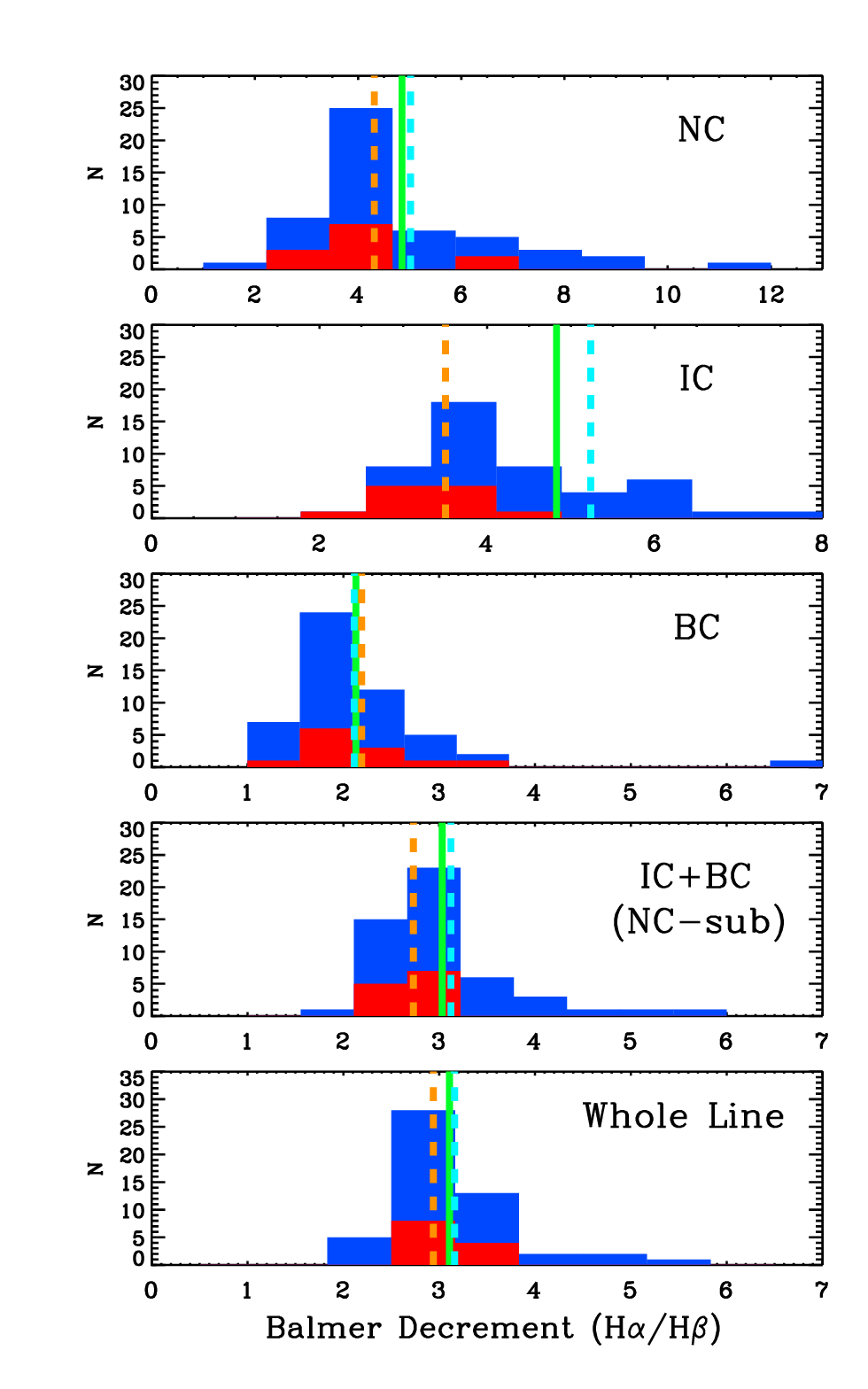}
\caption{Balmer decrement distributions of different Balmer line components.
In each panel the entire histogram shows the distribution
of the whole sample, with green solid line indicating the mean decrement value.
Red region highlights the distribution of the 12 NLS1s and the orange dashed
line indicates their mean decrement value. The cyan dashed line shows the mean
decrement value of the BLS1s.
The mean decrements are also listed in Table~\ref{balmerdec:hist:table}.}
\label{balmerdec:hist:plot}
\end{figure}

In addition to the luminosity and EW of Balmer line IC and BC, we can also
investigate the intrinsic properties of ILR and BLR by studying the profiles
of H$\alpha$ and H$\beta$.
{\bf The Balmer decrement is one of the main parameters to investigate.
It was reported that the change in 
Balmer decrement may arise from changes in the
physical conditions of the partially ionized line emitting
regions (Kwan \& Krolik 1979; Kwan \& Krolik 1981;
Mathews, Blumenthal \& Grandi 1980 and Canfield \& Puetter 1981).
For example, a decrease of Balmer decrement may
be due to an increase of electron density $Ne^{-}$, or an increase
of ionization parameter $\Xi$ (i.e. the ratio of the photon density
to the gas density), or an increase of Ly$\alpha$ optical depth
$\tau_{Ly\alpha}$ (Krolik \& McKee 1978; Davidson \& Netzer 1979; Shuder82).
A high Balmer decrement can also be explained
by a high dust abundance. This argument was used as an evidence in
Zhu et al. (2009) to support the link between ILR and dusty torus.
Therefore, Balmer decrements can be used as a clue to infer
the physical conditions of the emission line region, and so it is important
to obtain an accurate measurement of the decrement value.}
We calculated the Balmer decrement between H$\alpha$ 
and H$\beta$ for each line component. Figure~\ref{balmerdec:hist:plot}
shows our results. The five histograms from top to bottom show Balmer decrement 
(H$\alpha$/H$\beta$) distributions of NC, IC, BC, IC+BC and the whole 
line. The mean Balmer decrements with 1 standard
deviation are also listed in Table~\ref{balmerdec:hist:table}.

\begin{table}
 \centering
   \caption{The mean Balmer Decrements of different line components for NLS1 population, 
   BLS1 population and the whole sample.}
    \begin{tabular}{@{}cccc@{}}
    \hline
    Line Comp. & NLS1 & BLS1 & Whole Sample\\
    \hline
    NC &4.31$\pm$1.09&5.02$\pm$1.96&4.85$\pm$1.81\\
    \hline
    IC &3.50$\pm$0.61&5.24$\pm$2.33&4.83$\pm$2.18\\
    \hline
    BC &2.19$\pm$0.52&2.11$\pm$0.91&2.13$\pm$0.84\\
    \hline
    IC+BC &&&\\
    (i.e. NC-sub)&2.73$\pm$0.30&3.12$\pm$0.79&3.03$\pm$0.73\\
    \hline
    Whole Line &2.94$\pm$0.30&3.16$\pm$0.72&3.11$\pm$0.65\\
    \hline
    \end{tabular}
    \label{balmerdec:hist:table}
\end{table}

The results show that although the Balmer decrements of the whole
Balmer line distribute around 3, the situation for different line
components is quite different.  The NLR has a big range of Balmer
decrement values, with most probable value lying between 3.5-4.5, and
a mean value of 4.85$\pm$1.81. Therefore, it may imply the presence
of some dust in the NLR. The IC from ILR also has a big mean decrement
of 4.83$\pm$2.18, while The BC from BLR has a small mean decrement of
2.13$\pm$0.84. However, the Balmer decrement of the IC+BC distributes
around 3, which seems that the low decrement in BC and high decrement
in IC are artificially due to our multi-Gaussian decomposition. To
investigate this issue, we conduct the following study.

First, we think that the Balmer decrement we found for NLR is correct,
based on the fact that this component has a narrow width, and is
matched to the observed [OIII] profile.  For BLS1s, the profile of the
NC is easy to define as a small spike superposed on the much broader
line wings (see the Balmer line profiles in Paper-I); while for NLS1s,
untangling of the NC may introduce bigger uncertainties.  However, since
the NLR is far from AGN's centre and thus much less sensitive to the AGN's
central properties, we should not expect big difference in NLR's
properties between BLS1s and NLS1s, such as Balmer decrement.  This is
consistent with what we see in Figure~\ref{balmerdec:hist:plot}, that
the mean decrement values are similar between NLS1 and BLS1.  This
supports our assertion that the NC decomposition is reliable, and the
derived high Balmer decrement in NLR is real for the majority of
sample sources.

Second, we have reason to conclude that our decrement
distributions for the IC and BC are also intrinsic to Balmer line regions.
To prove this, we must first note that our combined H$\beta$ and H$\alpha$
line fitting has ensured very 
similar line decompositions for the two Balmer lines. To be specific, the IC and BC 
have the same central velocity shifts and FWHM in both H$\beta$ and H$\alpha$, 
but the IC and BC can have very different relative fluxes
(for more detailed description see Paper-I).
The observed decrement distribution differences between
IC and BC reveal real changes in the
decrement values across the emission line profile.

\begin{figure}
\centering
\includegraphics[bb=80 144 414 468, scale=0.75, clip=1]{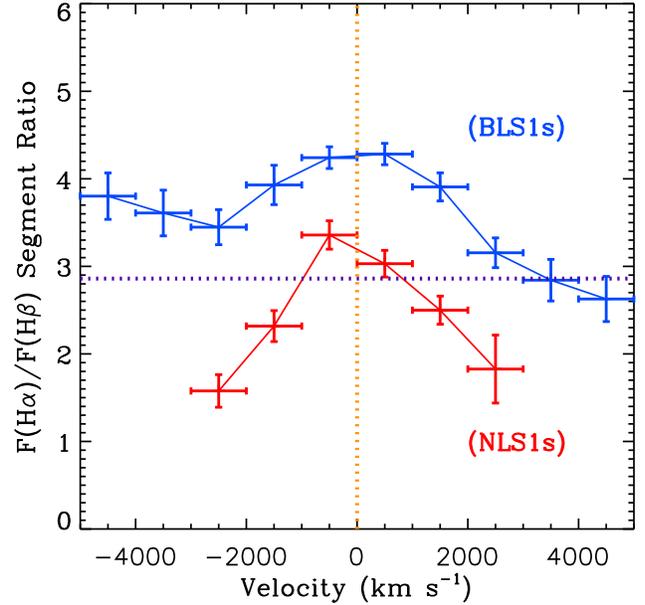}
\caption{Balmer decrement changing across the Balmer line profile 
from $+~5000~km~s^{-1}$ to $-~5000~km~s^{-1}$.
Each data point represents the average decrement
value in that segment with the vertical bar showing the $\pm$1 standard error. 
Blue points show the results for the BLS1s; red points show the results for NLS1s.
But due to the small line 
width of NLS1, the flux outside $+/-$ 3000 km s$^{-1}$ for NLS1s is of low S/N, 
thus only the mean decrement values in the central $+/-$ 3000 km s$^{-1}$ region 
were calculated and shown. The horizontal purple dotted line is a reference line
at F($\alpha$)/F($\beta$)~=~3.}
\label{balmerdec:segment:ratio:plot}
\end{figure}

To see this directly, we divided each of the two
Balmer lines into 10 segments in the velocity space
from -5000 km s$^{-1}$ to 5000 km s$^{-1}$, and calculated the decrement value 
in each segment. We performed this spectral analysis for each object in our sample. 
Then the average decrement value in each segment was calculated for sources 
with a reliable decrement measurement in that segment. This method is model-independent 
except for the subtraction of the local underlying continuum and removal of the 
[NII] $\lambda$6584,6550 doublets which uses the line fitting results from Paper-I.
The mean values for NLS1 subset and BLS1 subset in each segment
are plotted in Figure~\ref{balmerdec:segment:ratio:plot}. We see that the decrement 
peaks at the line centre and then decreases towards both sides.
This suggests a low decrement in the broad wings, which is mainly modelled by the BC, 
and so supports the low decrement value found for the BC.
For BLS1 subset, the decrement in the red side (indicating a positive velocity)
is lower than in the blue side (negative velocity).
It also appears in Figure~\ref{balmerdec:segment:ratio:plot} that NLS1s tend 
to have lower Balmer decrements than BLS1s in each segment.
However, Figure~\ref{balmerdec:hist:plot} shows
that the BC decrement distributions for the NLS1s and BLS1s are similar,
although they are sometimes of low contrast to the continuum.
Thus it suggests that the observed lower Balmer decrement of NLS1s
is mainly due to the lower decrement value of the IC in NLS1s, which is
also consistent with the IC decrement distributions in Figure~\ref{balmerdec:hist:plot}.
Therefore, we conclude that the observed differences in Balmer decrement 
distributions between different line components are mainly due to the
complex decrement status across the Balmer line profile,
with the broad wings having a lower decrement
value and contributing mainly to the BC.

There is also support for our findings from previous work by Shuder (1982)
(hereafter: Shuder82), who showed ratios of the H$\alpha$ and H$\beta$ lines for 18
Seyfert 1 AGNs. They reported that the average
H$\alpha$/H$\beta$ ratio ranged from 4.8 in the core to 2.2 in the wings,
which is very similar to what we have found.
They explained this as approaching the inner region of AGN, the
velocity dispersion would increase, and so do the $Ne^{-}$ and $\Xi$.
This is also consistent with the systematic inflow velocity
we found for the BC as we will discuss later.
Zhu09 also reported that the Balmer decrement in the IC was 4.78 and in BC it was 2.54,
although their line decomposition for H$\alpha$ and H$\beta$ were not linked 
as in our study. However, they explained this high decrement in IC
as due to the higher dust reddening in the ILR,
and so could support the link between ILR and dusty torus.
Therefore, the Balmer decrement change can be explained by
either dust abundance or line optical depth processes.
Since it is likely that BLR is closer to the core region than the ILR,
timing analysis such as detailed
reverberation mapping of Balmer lines can be used to distinguish the
radii between ILR and BLR. If the ILR is connected with BLR,
then the changes of physical parameters such $Ne^{-}$, $\Xi$ and
$\tau_{Ly\alpha}$ may be the explanation of the higher Balmer decrement
in ILR than in BLR.
Alternatively, if the ILR is confirmed to be a distinct region from BLR and
is close to the dusty torus, then dust reddening may be a more plausible answer.
Finally, we find no correlation between the Balmer decrement and the
bolometric luminosity, as was suspected by Shuder82.

\subsection{Balmer Line Component Fraction}
Zhu09 proposed an evolutionary scenario for the emission line region, which 
claimed that as the black hole mass and luminosity increased, ILR and BLR 
would become closer to each other and finally merge. In our sample 
49 of the 51 sources show the necessity of including two broad Gaussian
components to fit their Balmer lines,
as suggested by the Bayesian Information Criteria (BIC).
The fraction of each Gaussian component is calculated for both H$\alpha$ and H$\beta$,
and cross-correlated against the black hole mass, bolometric luminosity 
and the component line width. The results are listed in 
Table~\ref{balmer:line:fraction}. The dominant component in H$\beta$ is IC, 
but in H$\alpha$ it is BC. The NC fraction is $\sim$10\% in both lines.

We confirm a weak anti-correlation between the IC fraction and IC FWHM of
H$\beta$ (Spearman's $\rho_{s}=-0.4$, $d_{s}=-2$), similar to that found by Zhu09.
But no such anti-correlation was found in H$\alpha$.
So we conclude that it is still not clear whether the anti-correlation 
between the IC fraction and IC FWHM is an intrinsic property of the ILR.
We did not find any correlation in the IC or BC fractions vs. black hole mass
or bolometric luminosity. Therefore, the scenario proposed by Zhu09 regarding the 
geometry of ILR and BLR can not be confirmed by the results of our study.
Instead, our results suggest that the ILR may simply
be an intermediate region between BLR and NLR regardless of the black
hole mass and bolometric luminosity. As the black hole mass increases, the luminosities
of NC, IC and BC in both H$\alpha$ and H$\beta$ all increase,
but the luminosity of IC and BC increase more significantly than the NC, resulting
in the significant anti-correlation between the black hole mass and NC fraction.

\begin{table}
 \centering
   \caption{The fraction of each Balmer line component and its correlations 
   with black hole mass (M$_{BH}$), bolometric luminosity (L$_{bol}$) and the 
   H$\beta$ IC+BC FWHM. $\rho_{s}$ and $d_{s}$ are Spearman's rank coefficients 
   as explained in Table~\ref{app:table:spearman}.}
    \begin{tabular}{@{}cccccccc@{}}
    \hline
    & {\it frac\/}&\multicolumn{2}{c}{vs. M$_{BH}$} & \multicolumn{2}{c}{vs. L$_{bol}$} & \multicolumn{2}{c}{vs. FWHM}\\
    H$\alpha$ & &$\rho_{s}$&$d_{s}$&$\rho_{s}$&$d_{s}$&$\rho_{s}$&$d_{s}$\\
    \hline
    NC & 10$\pm$8  &-0.6 & -6 &-0.2 & -1 &-0.7 & -7\\
    IC & 53$\pm$11 & 0.1 & -0 &-0.1 & -0 & 0.1 & -0\\
    BC & 37$\pm$11 & 0.4 & -2 & 0.3 & -2 & 0.4 & -2\\
    IC+BC&90$\pm$8 & 0.6 & -6 & 0.2 & -1 & 0.7 & -7\\
    \hline
    H$\beta$ & {\it \% \/} &$\rho_{s}$&$d_{s}$&$\rho_{s}$&$d_{s}$&$\rho_{s}$&$d_{s}$\\
    \hline
    NC & 7$\pm$6   &-0.6 & -5 &-0.2 & -1 &-0.5 & -4\\
    IC & 38$\pm$12 &-0.0 & -0 & 0.0 & -0 &-0.4 & -2\\
    BC & 55$\pm$14 & 0.4 & -2 & 0.1 & -0 & 0.5 & -5\\
    IC+BC&93$\pm$6 & 0.6 & -5 & 0.2 & -1 & 0.5 & -4\\
    \hline
    \end{tabular}
    \label{balmer:line:fraction}
\end{table}

\subsection{Balmer Line Shape}
\label{balmer:line:shape}
\subsubsection{Dependences on Eddington ratio}
The complex Balmer line profile can often be well modelled using contributions
from the NLR, ILR and BLR, but it is also likely that local turbulence
in the BLR may further broaden the Balmer line profile and be responsible
for the presence of a very broad wing. Collin et al. (2006)
(hereafter: Collin06) divided the reverberation-mapped sample into two 
populations: sources in the first population have narrower H$\beta$ lines
with more extended wings, along with higher Eddington ratios; while sources 
in the second population have broader but flat-topped H$\beta$ lines,
together with lower Eddington ratios. They found a weak anti-correlation 
between FWHM/$\sigma_{line}$ and Eddington ratio where $\sigma_{line}$ 
is the second moment of H$\beta$ line (see the definition given in
Peterson et al. 2004). This can be explained in terms of higher turbulence
in the core region of high Eddington ratio AGNs. We can explore this
result for our sample.

The second moment was measured from the NC subtracted H$\beta$ line profile.
It is clear from the definition that the flux farther from the line centre
would have more contribution to the total second moment, thus
$\sigma_{line}$/FWHM$_{IC+BC}$ can be a representative of the broad wing
strength compared to the whole line profile, i.e. a higher value of
$\sigma_{line}$/FWHM$_{IC+BC}$ corresponds to stronger broad wings.
In our Balmer line profile fitting, the broad wing is mainly modelled by
the BC, thus the FWHM ratio between BC and IC should have similar
physical meaning as $\sigma_{line}$/FWHM$_{IC+BC}$. We correlate both 
$\sigma_{line}/$FWHM$_{IC+BC}$ and FWHM$_{BC}/$FWHM$_{IC}$ with Eddington 
ratio (L$_{bol}$/L$_{Edd}$).

\begin{figure}
\centering
\includegraphics[bb=60 72 396 648, scale=0.7, clip=1]{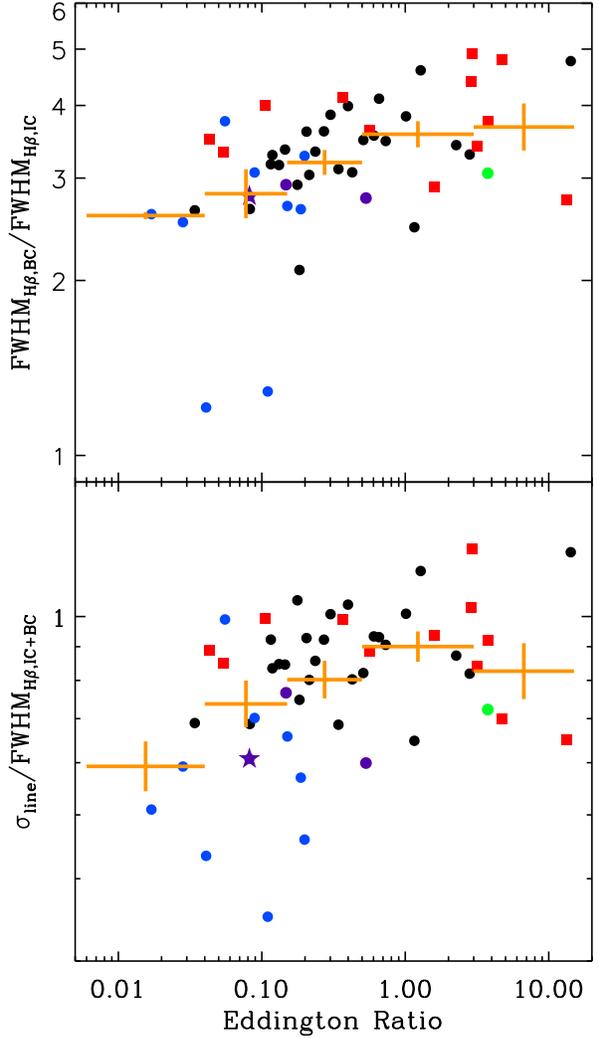}
\caption{The H$\beta$ line shape correlation with Eddington ratio.
The upper panel uses FWHM$_{BC}/$FWHM$_{IC}$ to represent H$\beta$ shape,
while the lower panel uses $\sigma_{line}$/FWHM$_{IC+BC}$ instead.
In each panel the various symbols represent the same type of sources as
in Figure~\ref{hblum:210lum:plot}. The orange data points are the binned
data for different Eddington ratio bins with 1 standard error on the Y-axis.}
\label{linewidth:eddr:plot}
\end{figure}

Figure~\ref{linewidth:eddr:plot} shows our 
results. Orange points are the binned values over X-axis (Eddington ratio) 
with one standard error in the Y-axis (line width ratio).
{\bf The errors in Eddington ratio and line width ratio
for each data point are dominated by systematical uncertainties
in the spectral fitting which is difficult to quantify (see Paper-I),
therefore individual error-bars are not shown.
Such uncertainties will increase the dispersion in this correlation plot.}
However, positive correlations have been confirmed
in both panels with Spearman's rank test 
being $\rho_{s}$=0.47, $d_{s}=10^{-3}$ (upper panel) and $\rho_{s}$=0.35, 
$d_{s}=10^{-2}$ (lower panel). 
Our results confirm the correlation between H$\beta$ line profile and 
L$_{bol}$/L$_{Edd}$. Interestingly, we also find similar correlations 
when replacing L$_{bol}$/L$_{Edd}$ with $\kappa_{2-10keV}$ and $\alpha_{ox}$. 
For $\sigma_{line}$/FWHM$_{IC+BC}$, the Spearman's test has $\rho_{s}$=0.46 
($d_{s}=10^{-3}$) for $\kappa_{2-10keV}$ and $\rho_{s}$=0.51 ($d_{s}=10^{-4}$) for 
$\alpha_{ox}$. For FWHM$_{BC}/$FWHM$_{IC}$, the Spearman's test has 
$\rho_{s}$=0.54 ($d_{s}=10^{-4}$) for $\kappa_{2-10keV}$ and $\rho_{s}$=0.54 
($d_{s}=10^{-4}$) for $\alpha_{ox}$. But since the Eddington ratio is an  
intrinsic AGN parameter, and it also correlates with both $\kappa_{2-10keV}$ 
and $\alpha_{ox}$ {\bf (Vasudevan \& Fabian 2007; Vasudevan \& Fabian 2009;
Lusso et al. 2010; Grupe et al. 2010),} it may be the driving parameter that regulates
the Balmer line shape, as was also suggested by Collin06. 
Therefore, the FWHM of the BC may depend primarily on black hole mass, 
but it may also be regulated by L$_{bol}$/L$_{Edd}$. This also explains
the stronger correlation between black hole mass and the IC FWHM than
between black hole mass and the BC FWHM.

\subsubsection{Inflow Implied by the Balmer Line Profile}

Another effect that may change the width of the whole Balmer line is 
the systematic velocity structure of the BLR as evinced by the 
general redshift of the BC and IC. It was found 
previously that both the IC and BC may be associated with inflows 
(e.g. Sulentic et al. 2000; Hu et al. 2008). Our analysis also shows that
for the whole sample both IC and BC have a wide range of velocity shifts
relative to the central component of [OIII] $\lambda$5007, but on average
we find a statistically significant shifts of 100 km s$^{-1}$
for the IC and 550 km s$^{-1}$ for the BC. 
The inflow velocity we find for the BC is also consistent with
the $\sim$400 km s$^{-1}$ typical inflow velocity of the FeII emission
features found by Hu et al. (2008), supporting their conclusion that
FeII emitting region may trace some portions of the BLR exhibiting inflow.
It seems probable that there is a velocity gradient within the BLR clouds,
with the inner region of the BLR having a higher inflow speed,
which gives rise to the extended red wing. However, the multi-Gaussian 
Balmer line decomposition method used in this study cannot resolve
the detailed changes in the kinematics and physical conditions within BLR.
Therefore, a much more detailed broad line spectral and timing study is required.
The relative velocity shifts between the IC and BC determines the asymmetry of
the Balmer lines, but we did not find any significant correlation between the
relative velocity shift and Eddington ratio. So the relative
velocity shift cannot explain the correlation between
$\sigma_{line}/FWHM_{IC+BC}$ and L$_{bol}$/L$_{Edd}$.

\begin{figure*}
\centering
\includegraphics[scale=0.85, clip=1, angle=90]{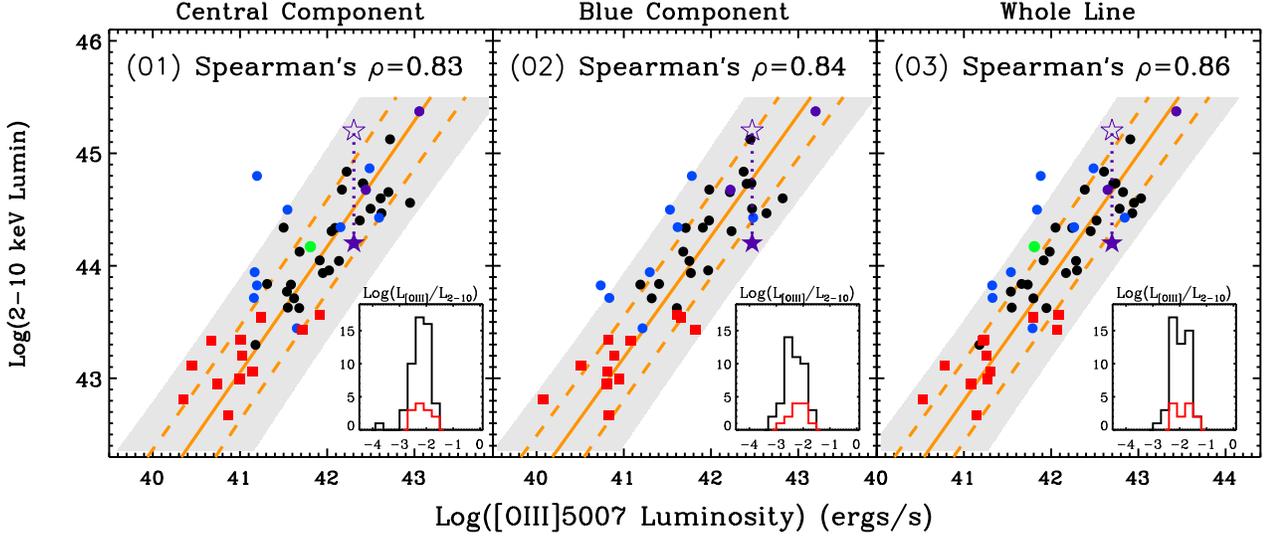}
\caption{The luminosity correlations between components of 
[OIII] $\lambda$5007 and L$_{2-10keV}$. All symbols and lines 
have the same meanings as in Figure~\ref{hblum:210lum:plot}. In each panel, a histogram is shown for
the Log(L$_{[OIII]\lambda5007}$/L$_{2-10keV}$) values of our sample, with the red histogram
highlighting the NLS1s.}
\label{o3:210kev:plot}
\end{figure*}

\subsection{Summary of results for the ILR and BLR}
To summarize the previous subsections, we propose the following 
characteristics of the ILR and BLR. First, these two regions are likely to
be closely related, with the ILR being an extension of the BLR.
We found no evidence to support the ILR to be a distinct region from BLR.
The inner region of the BLR may produce the red wing of the Balmer profile,
indicating an systematic inflow velocity. The physical parameters
change continuously from the ILR into the BLR, probably with increasing
electron density, ionization parameter and Ly$\alpha$ optical depth.
The inflow velocity of the ILR gas is smaller than the BLR.
The possibility that the ILR is associated with the dusty torus cannot
be ruled out, and indeed its Balmer Decrement is higher than found for the BLR.
Unfortunately we are unable to draw any firm
conclusions about the geometry of the ILR or BLR.
A second-order factor such as the covering factor of the ILR and BLR may cause
the faster than linear dependence of the Balmer IC and BC luminosities
on the continuum luminosity. Considering the tight
correlations between the IC FWHM, BC FWHM and black hole mass, 
both the ILR and BLR should be gravitational bound and Virialized.
For the BC from BLR, its FWHM may further be affected by the Eddington
ratio through the process of local turbulence. Different inflow velocities
of the ILR and BLR may also modify the shape of the Balmer lines.

\begin{figure*}
\centering
\includegraphics[bb=54 360 558 800, scale=0.8, clip=1]{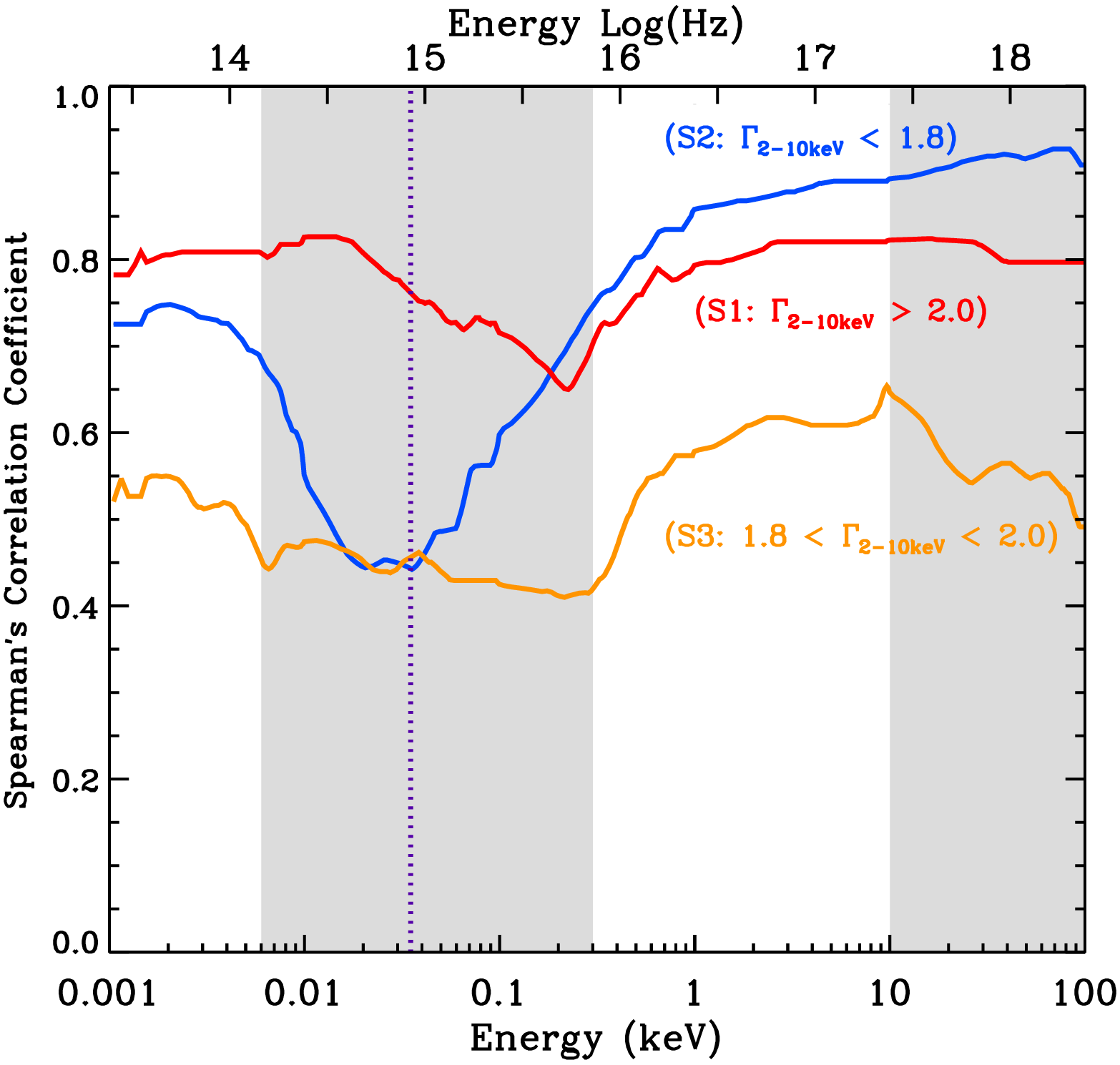}
\caption{The `SED to [OIII] $\lambda$5007 Correlation Spectra (SOCS)'.
This is produced by calculating the Spearman's rank coefficient between the 
[OIII] $\lambda$5007 luminosity and the luminosity contained in each energy 
bin of broadband SED, thus the bigger coefficient indicates the better correlation in
that energy bin. Lines of different color show the SOCS of different subsets
as been labelled in the plot. The $\Gamma_{2-10keV}{\ge}2.0$ subset (S1: red line)
contains 16 AGNs; the $\Gamma_{2-10keV}{\le}1.8$ subset contains
(S2: blue line) 18 AGNs; the $1.8<\Gamma_{2-10keV}<2.0$ subset (S3: orange line)
contains 16 AGNs.
Only spectral ranges below 0.006 keV and above 0.3 keV have observational data. 
The ionizing flux responsible for [OIII] $\lambda$5007 emission is
above 0.035 keV as shown by the purple dotted line. The two shaded regions
are where model extrapolation was used.}
\label{o3xraycroscor}
\end{figure*}

\section{Properties of Emission Line [OIII] $\lambda$5007}
\label{section:OIII}

The NLR may extend hundreds of parsecs from the AGN's compact core. It
is ionized by the central continuum in a bi-conical geometry
with an axis defined by the plane of the dusty torus. 
Since the NLR extends far from the dusty torus, 
the intrinsic dust reddening is expected to be low. 
Therefore in both Type 1 and Type 2 AGNs, the luminosity of
narrow optical emission lines from NLR can provide an
orientation-independent estimate of the central ionizing radiation
(e.g. Mulchaey et al. 1994; Heckman 1995).

\subsection{[OIII] $\lambda$5007 Component Luminosity vs. L$_{2-10keV}$}

As one of the strongest narrow forbidden lines, [OIII] $\lambda$5007 
is often employed as a proxy to estimate the intrinsic 
luminosity of type 2 AGN (e.g. Heckman et al. 2004; Brinchmann et al. 2004).
This is not only because of its large EW, 
but also because it is free from serious contamination of other spectral features. 
Heckman et al. (2005) (hereafter: Heckman05) used [OIII] $\lambda$5007 as 
an optical selection criteria to study the difference between 
optical and X-ray selected AGN samples. They showed a tight 
correlation between [OIII] $\lambda$5007 luminosity and hard 
X-ray luminosity for Type 1 AGNs in both optical and X-ray 
selected samples. But the correlation is much weaker for 
an optically selected Type 2 AGN sample, which is mainly due to 
their X-ray weakness resulting from intrinsic photoelectric 
absorption. Since our Type 1 AGN sample has been carefully
selected based on the high quality optical and X-ray
spectra and the absence of severe absorption,
our [OIII] $\lambda$5007 vs. rest frame 2-10 keV luminosity correlation 
should be indicative of intrinsic connections. We also find that 
the line profile of [OIII] $\lambda$5007 consists of two components, a 
dominant central component and a blue-shifted component.
Thus we analyze these two components separately.

Figure~\ref{o3:210kev:plot} shows our results. We find that 
both components in [OIII] $\lambda$5007 have very strong  
correlations with L$_{2-10keV}$, as confirmed by the Spearman 
rank test. The total [OIII] $\lambda$5007 line correlates 
slightly better with L$_{2-10keV}$ than the two separate 
components, suggesting that the central ionizing radiation 
ionizes both the outflowing [OIII] $\lambda$5007 region 
(which produces the blue component), and the spatially more
extended [OIII] $\lambda$5007 region (which produces the central component). 
The solid orange line is the regression line assuming L$_{2-10keV}$
to be the independent variable.
The two dashed orange line indicate the $\pm$1$\sigma$ region for new observation.
The shaded region denotes the $\pm$2$\sigma$ region.

Again, we put our two excluded objects on the correlation. Mrk 110
(green circle) sits on the best fit line as neither its X-ray nor its
[OIII] line luminosity are affected by the optical continuum
variability. The X-ray weakness of PG 1004+130 (purple star) does not
cause strong deviation in these correlation plots considering the dispersion.
The other outlier is 1RXS J122019 (the blue circle point farthest
from the shaded region) whose SDSS spectrum shows that this is
one of the `broadest' BLS1s, but its narrow lines
(including [OIII] $\lambda$5007,4959 doublets) are much
weaker relative to the broad lines than any other sources in our sample.
It is possible that compared with other sources,
1RXS J122019 has a smaller NLR covering factor.

We also calculated Log(L$_{[OIII]{\lambda}5007}$/L$_{2-10keV}$) using
the luminosity of the whole [OIII] $\lambda$5007 line. We derived a
mean value of -1.88$\pm$0.31 for the whole sample, and -1.78$\pm$0.30
for the 12 NLS1s.  This is consistent with but slightly lower than
-1.59$\pm$0.48 reported by Heckman05 based on their sample of 20
Seyfert 1s, which is likely due to the fact that their sample includes
sources with strong X-ray absorption, e.g. they included Type 1 AGNs
such as NGC 3227 and Mrk 766 whose X-ray spectra are absorbed by warm
absorber.  Trouille \& Barger (2010) (hereafter: TB10) reported
-1.85$\pm$0.5 for their 19 BLAGN sources with $z<$ 0.5 and
-1.76$\pm$0.5 for their $\sim$100 BLAGNs with $z<$ 0.85.
Georgantopoulos \& Akylas (2010) (hereafter: GA10) reported
-1.98$\pm$0.39 for their 34 Seyfert 1s.  These values are all
consistent with ours within 1$\sigma$.  Our mean
Log(L$_{[OIII]{\lambda}5007}$/L$_{3-20keV}$) value is -2.00$\pm$0.33,
which is consistent with -1.96$\sim$2.14 reported by Heckman05.  Note
that compared with previous works, our sample is spectrally `cleaner',
i.e. more carefully selected for low reddening and absorption sources.
Thus our mean Log(L$_{[OIII]{\lambda}5007}$/L$_{2-10keV}$) and
Log(L$_{[OIII]{\lambda}5007}$/L$_{3-20keV}$) are better constrained as
having smaller dispersions.

The correlations between [OIII] $\lambda$5007 and L$_{2-10keV}$
can be used to estimate the intrinsic hard X-ray luminosity especially
for nearby Type 2 AGNs and calculate the X-ray luminosity function
(e.g. Sazonov \& Revnitsev 2004; Shinozaki et al. 2006;
Yencho et al. 2009; GA10).
We present our OLS bisector regression lines below:\\
(i) L$_{[OIII]{\lambda}5007}$) expressed by $L_{2-10keV}$ and $L_{3-20keV}$:
\begin{eqnarray}
\label{o3lum:210lum:eqn:1}
Log L_{[OIII]5007}=(1.06{\pm}0.05)Log L_{2-10} - (4.44{\pm}2.64)\\
Log L_{[OIII]5007}=(1.02{\pm}0.05)Log L_{3-20} - (2.93{\pm}2.65)&
\end{eqnarray}
(ii) $L_{2-10keV}$ and $L_{3-20keV}$ expressed by L$_{[OIII]{\lambda}5007}$):
\begin{eqnarray}
\label{o3lum:210lum:eqn:2}
Log L_{2-10}=(0.94{\pm}0.05)Log L_{[OIII]5007} + (4.20{\pm}2.26)\\
Log L_{3-20}=(0.98{\pm}0.05)Log L_{[OIII]5007} + (2.87{\pm}2.43)&
\end{eqnarray}
Note that since the luminosity measurements of
both [OIII] $\lambda$5007 and L$_{2-10keV}$ may contain uncertainties from the
intrinsic absorptions and variability, the OLS bisector regression method is more
appropriate than the standard OLS method used in previous works.
Combining previous results from Heckman05, TB10 and GA10 with ours, we can conclude
that such tight luminosity correlations between hard X-ray and [OIII] appear valid
for at least $z<$ 0.85, Log(L$[OIII]{\lambda}5007$)=38$\sim$44 and
Log(L$_{2-10keV}$)=40$\sim$46.

\subsection{The SED to [OIII] $\lambda$5007 Correlation Spectra (SOCS) }
\label{section:SOCS}
The ionizing energy for [OIII] $\lambda$5007 is 0.035 keV, thus all
photons above this energy can in principle produce [OIII]
$\lambda$5007 emission.  As another application of CST, we
cross-correlate [OIII] $\lambda$5007 luminosity with the luminosity
contained in the continuum SED in each energy bin from optical to
X-ray, and produce the `SED to [OIII] $\lambda$5007 correlation
spectrum (SOCS)'.  Our broadband SED model consists of three
components: the disc, soft X-ray Comptonisation and hard X-ray
Comptonisation. The soft and hard Comptonisation components can both
contain a significant amount of energy above 0.035 keV, the disc
emission may also extend above this energy when the black hole mass is
small and mass accretion rate is high (Done et al. 2011).  We divided
our sample into three subsets: $\Gamma_{2-10keV}{\ge}2.0$ (S1: 16
AGNs), $\Gamma_{2-10keV}{\le}1.8$ (S2: 18 AGNs) and
$1.8<\Gamma_{2-10keV}<2.0$ (S3: 16 AGNs).  PG 1004+130 is excluded due
to its unique SED shape but Mrk 110 is now included since the previous
section shows that its [OIII] and hard X-ray luminosity are not
distorted by the optical continuum variability. S1 includes all 12
NLS1s, while S3 contains the broadest BLS1s. A SOCS was calculated for
each of the three subsets.

Figure~\ref{o3xraycroscor} shows the resultant SOCSs.
Note: in this study only spectral ranges below 0.006 keV and above 0.3 keV
have observational data. Overall, [OIII] $\lambda$5007
correlates best with hard X-rays above 2 keV. It is also well-correlated
with the optical emission. However, the correlation in the UV/X-ray region is poor,
which may be caused by the spectral modification due to Galactic and
intrinsic extinction. This indicates that the hard X-ray ionizing flux also
has a strong link with the optical flux which is presumably
dominated by accretion disc emission.

Regarding the SOCS of different subsets, we find that S1 has 
the strongest correlations in the optical/UV band, which implies that
our broadband SED fitting may be more reliable for sources in S1 whose
soft X-ray excess is more likely to be a real extra component
(Middleton et al. 2009; Jin et al. 2009; Middleton, Uttley \& Done).
The S2 group shows highly significant
correlations in the hard X-ray band, which may imply that the hard X-ray 
power law tail of $\Gamma_{2-10keV}{\le}1.8$ is an intrinsic separate
component rather than being an artifact caused by absorption or 
reflection (Done et al. 2011). Therefore, the S1 and S2 groups may indeed
represent two distinct types of AGNs (e.g. NLS1s and BLS1s). We also find that
the correlation for S3 in optical and X-ray bandpasses
is much less significant than for either S1 or S2.
This may indicate that other spectral factors
such as absorption and reflection, may be more important for the sources
in S3, in which case our three-component SED model is too simple to recover 
their intrinsic SEDs.

\subsection{Outflow of NLR Implied by [OIII] $\lambda$5007 Profile}

The two components in the profile of [OIII] $\lambda$5007 have been reported previously.
Bian, Yuan \& Zhao (2005) indicated that these two components
are related to two physically distinct regions. Komossa et al. (2008) also
reported blue outliers in [OIII] $\lambda$5007 whose blue-shift velocity
is up to 500$\sim$1000 km s$^{-1}$, favoring a decelerating wind NLR scenario.
These results are all confirmed by our study.
We find that for our sample, the velocity shift of the blue component
in [OIII] $\lambda$5007 relative to the central component ranges from
-610 to -0 km s$^{-1}$, and the mean velocity is $-130^{+80}_{-230}~km~s^{-1}$.
We also find a strong correlation between the FWHM and velocity
shift of the blue component, as Spearman rank test gives:
$\rho_{s}$=0.52 and $d_{s}=10^{-4}$. The larger FWHM of the blue
component implies a smaller distance from AGN's core region,
so this correlation suggests that the outflow velocity of inner NLR  
emitting [OIII] $\lambda$5007 is higher than that in the outer NLR. 
An outflow speed decreasing as it flows away from the centre is a
signature of decelerating wind.

\section{Summary and Conclusions}
In this paper, we made use of the detailed spectral fitting of an AGN sample reported
in Paper-I, to study their optical spectral properties using their hard
X-ray luminosity as a diagnostic. Our study focused on the H$\beta$,
H$\alpha$ and [OIII] $\lambda$5007 emission lines and the underlying continuum.
The main results are summarized below.

$\bullet$ The OXCSs have been constructed for different subsets of AGNs
using our new spectral analyzing technique called CST.
The OXCSs reveal many correlation features
with L$_{2-10keV}$ across the entire optical spectrum.
Some were known previously, others are new. For example,
the entire optical underlying continuum strongly correlates with L$_{2-10keV}$.
[NeIII] $\lambda\lambda$3869/3967, [OI] $\lambda\lambda$6300/6364, 
[OII] $\lambda\lambda$3726/3729, [OIII] $\lambda\lambda$4959/5007 and the IC and
BC in Balmer lines all well correlates with L$_{2-10keV}$ especially for BLS1s.
However, stellar absorption lines, FeII and the NC in the Balmer lines has much weaker
or no correlation with L$_{2-10keV}$. We find some evidences for differences
in the OXCSs between NLS1s and BLS1s.

$\bullet$ The cross-correlation between luminosities of H$\beta$ and H$\alpha$ 
line components and the broadband SED components were performed.
The results suggest that among the three SED components,
the hard X-ray power law component correlates the best with Balmer line luminosity,
and the correlations strengthen from the NC, IC to BC of Balmer lines. 
This supports the view that the BC has the closest link
with AGN's central UV/X-ray continuum emission.

$\bullet$ Significant correlations were found between
the H$\beta$ component EWs and L$_{2-10keV}$, $\kappa_{2-10keV}^{-1}$, H$\beta$ FWHM
and black hole mass, although these correlations become weaker for the BLS1 subset alone.
By cross-correlating Balmer line component EWs with L$_{5100}$,
no evidence for the `Baldwin Effect' was found for the IC and BC,
but such effect is weakly detected for the NC.

$\bullet$ Our results suggest a faster than linear dependence of
Balmer line IC and BC luminosities on the underlying continuum
(e.g. L$_{2-10keV}$ and L$_{5100}$; Equation~\ref{equ:exp-dep}),
implying the presence of a second-order factor.  We propose that this
second-order effect could be the covering factor of the BLR and ILR
seen by the central UV/X-ray continuum, so that higher L$_{2-10keV}$
and L$_{5100}$ sources may also have larger ILR and BLR covering
factors.

$\bullet$ We carried out detailed Balmer line shape studies in order to 
reveal the nature of ILR and BLR. We found that the Balmer Decrement value,
defined by H$\alpha$/H$\beta$,
peaks at the line centre and decreases towards both sides, with the red wing having 
a lower decrement than the blue wing for BLS1 subset. This was also consistent with IC's 
average decrement value of 4.83$\pm$2.18 compared to the BC's 2.13$\pm$0.84. 
These results, along with the systematic inflow speed we found in the BC
(mean velocity: 550 km s$^{-1}$),
support the scenario that the inner region of BLR forms the red wing while the 
outer edge links with the ILR. Compared to the ILR, the BLR may have higher inflow speed,
higher electron density, larger ionization parameter or higher Ly$\alpha$ optical depth, 
A weak correlation between the shape of Balmer line profile and Eddington ratio was
confirmed. A higher Eddington ratio corresponds to a more extended wing
relative to the overall Balmer line structure. This implies that the velocity width of
the Balmer line is not simply determined by the black hole mass, but also affected by
local turbulence whose strength depends on the Eddington ratio.
The higher Balmer decrement in ILR than in BLR could also be explained as ILR has higher
dusty abundance, but we found no other evidence to support ILR's link with the dusty torus.
A weak anti-correlation between the EW of Balmer line NC and black hole mass was found.

$\bullet$ In our study of [OIII] $\lambda$5007, we refined its tight correlation 
with L$_{2-10keV}$ and L$_{3-20keV}$.
We found that the blue and central components of [OIII] $\lambda$5007 should  
be added together to provide the best correlation with hard X-rays. Using our best-fit
broadband SEDs from Paper-I, we produced the SOCSs for different sample subsets.
The SOCSs show strong correlations between [OIII] $\lambda$5007 luminosity and the
continuum luminosities in either optical or hard X-ray bandpass.
Subset S1 and S2 both have highly significant correlations in the
hard X-ray band, which implies that the shape of hard X-ray power law tail in these 
two subsets are intrinsic in spite of their totally different photon indices.
But the SED of moderate $\Gamma_{2-10keV}$ sources in S3 may be more
complex.

$\bullet$ The mean outflow velocity of the blue component in
[OIII] $\lambda$5007 is $-130^{+80}_{-230}~km~s^{-1}$.
The strong correlation between the FWHM and velocity shift of the blue component 
in [OIII] $\lambda$5007 suggests that the outflow speed of [OIII] $\lambda$5007 
clouds decreases from the central region outwards, suggesting a decelerating wind.

$\bullet$ In this paper, we present well constrained equations which can be used  
to convert between the luminosity of Balmer line broad component and the 
intrinsic L$_{2-10keV}$ (Equation 1$\sim$4),
between the intrinsic L$_{5100}$ and L$_{2-10keV}$ (Equation 5$\sim$6),
and between the [OIII] $\lambda$5007
luminosity and the intrinsic L$_{2-10keV}$ and L$_{3-20keV}$ (Equation 8$\sim$11).
We suggest that these equations be used for inferring the intrinsic optical and X-ray
luminosities of obscured sources such as BAL quasars or Type 2 AGNs,
and for calculating the X-ray luminosity function. Considering
the limited redshift range of and size of our sample, similar studies should be carried
out on larger samples to test the robustness and evolution of these equations
at high redshift, which requires high quality infrared spectra.

\section*{Acknowledgements}
We are extremely grateful to Hermine Landt for her useful discussion and
suggestions on the draft of this manuscript.
We sincerely thank Dirk Grupe for his helpful comments and suggestions.
C. Jin acknowledges financial support through Durham Doctoral Fellowship.
This work is partially based on the data from SDSS,
whose funding is provided by the Alfred P. Sloan Foundation,
the Participating Institutions, the National Science Foundation,
the U.S. Department of Energy, the National Aeronautics and Space Administration,
the Japanese Monbukagakusho, the Max Planck Society,
and the Higher Education Funding Council for England.
This work is also partially based on observations obtained with XMM-Newton,
an ESA science mission with instruments and contributions directly
funded by ESA Member States and the USA (NASA).

\appendix
\onecolumn
\centering

\section{Supplement of Balmer Component Correlation Plots}

\begin{figure*}
\centering
\includegraphics[scale=0.65,clip=1,angle=90]{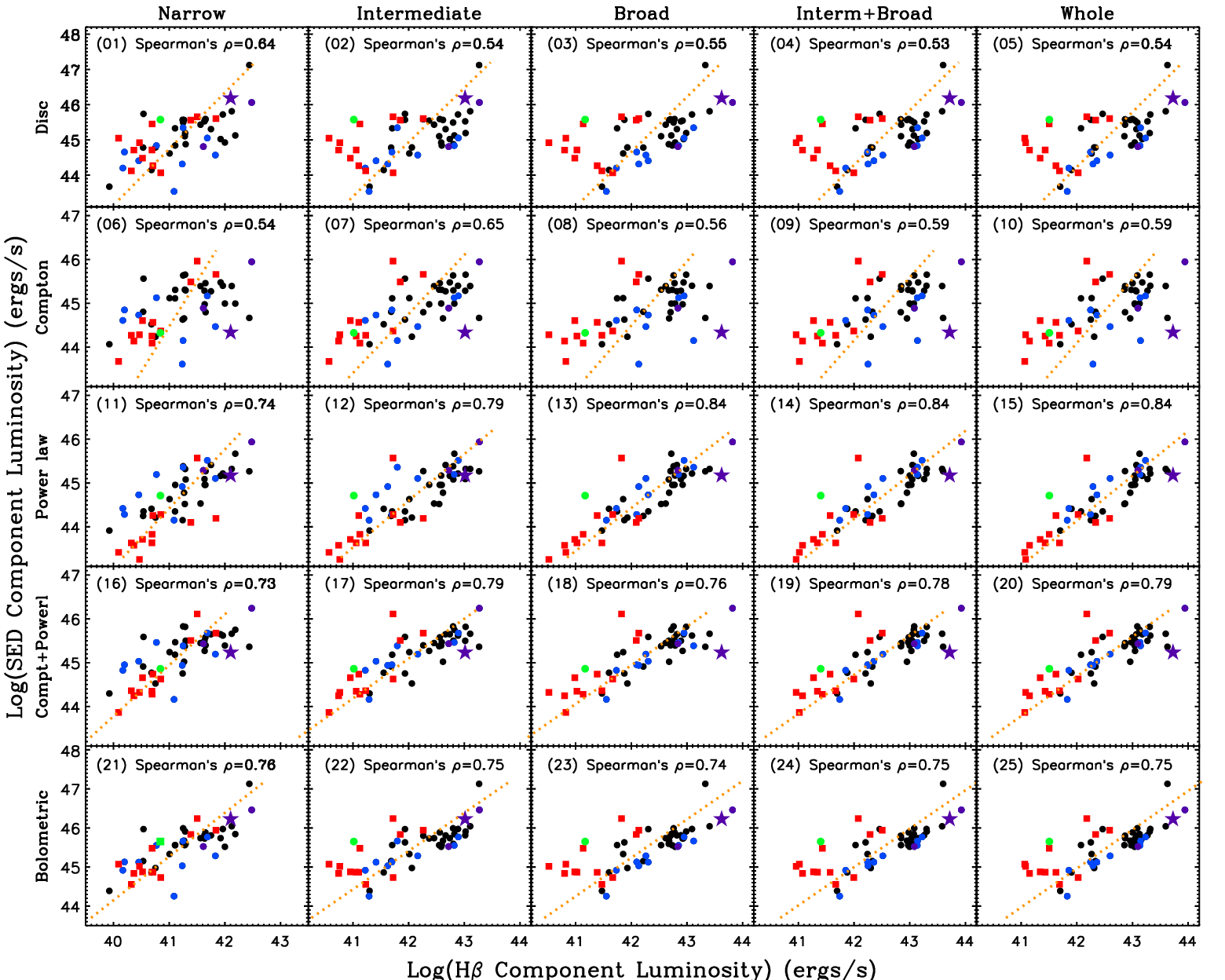}
\caption{The luminosity correlations between H$\beta$ line components and 
SED components. Red points represent NLS1s; blue points represent the 
broadest H$\beta$ line BLS1s; green point is Mrk 110; purple star is
PG 1004+130; purple symbols indicate radio loud sources. The orange
dotted line denotes the OLS regression line assuming the SED component
luminosity is the independent variable. Spearman's rank correction
coefficient $\rho_{s}$ for the whole sample is also given in each panel.}
\label{app:hblum:sedlum:plot}
\end{figure*}

\begin{figure*}
\centering
\includegraphics[scale=0.65,clip=1,angle=90]{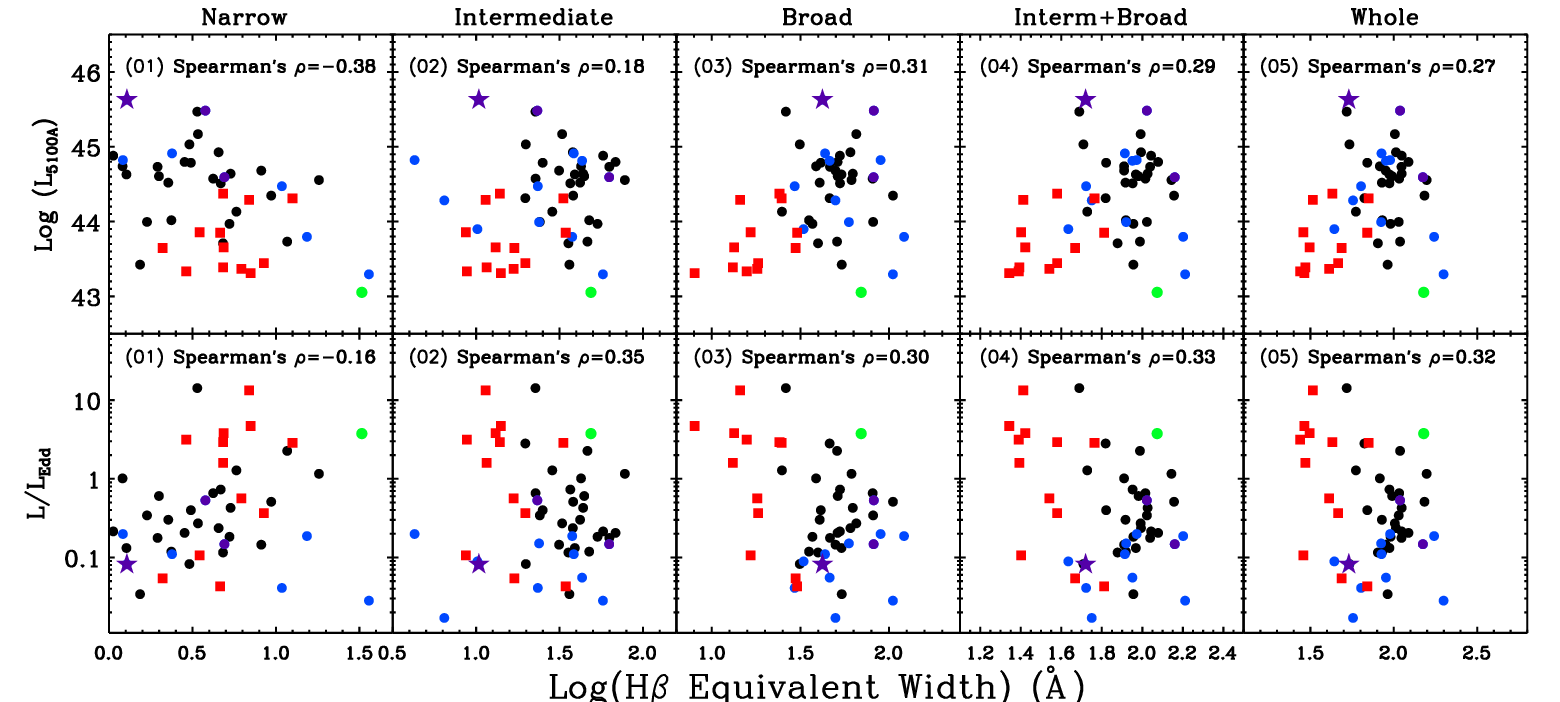}
\caption{The cross-correlation plots between H$\beta$ line component EWs and L$_{5100A}$ (the monochromatic luminosity at 5100{\AA}) and L/L$_{Edd}$ (the Eddington ratio) Different symbols have the same meaning as in Figure~\ref{app:hblum:sedlum:plot}. Spearman's $\rho$ is given in each panel.}
\label{app:fig:balmerew}
\end{figure*}

\clearpage

\section{The Spearman's Rank Correlation Matrix between H$\alpha$, H$\beta$, [OIII] $\lambda$5007 Line Components and SED Components}
\begin{table*}
    \begin{minipage}{180mm}
    \centering
        \begin{tabular}{c}
        \includegraphics[bb=20 200 596 1000, scale=0.87,clip=1]{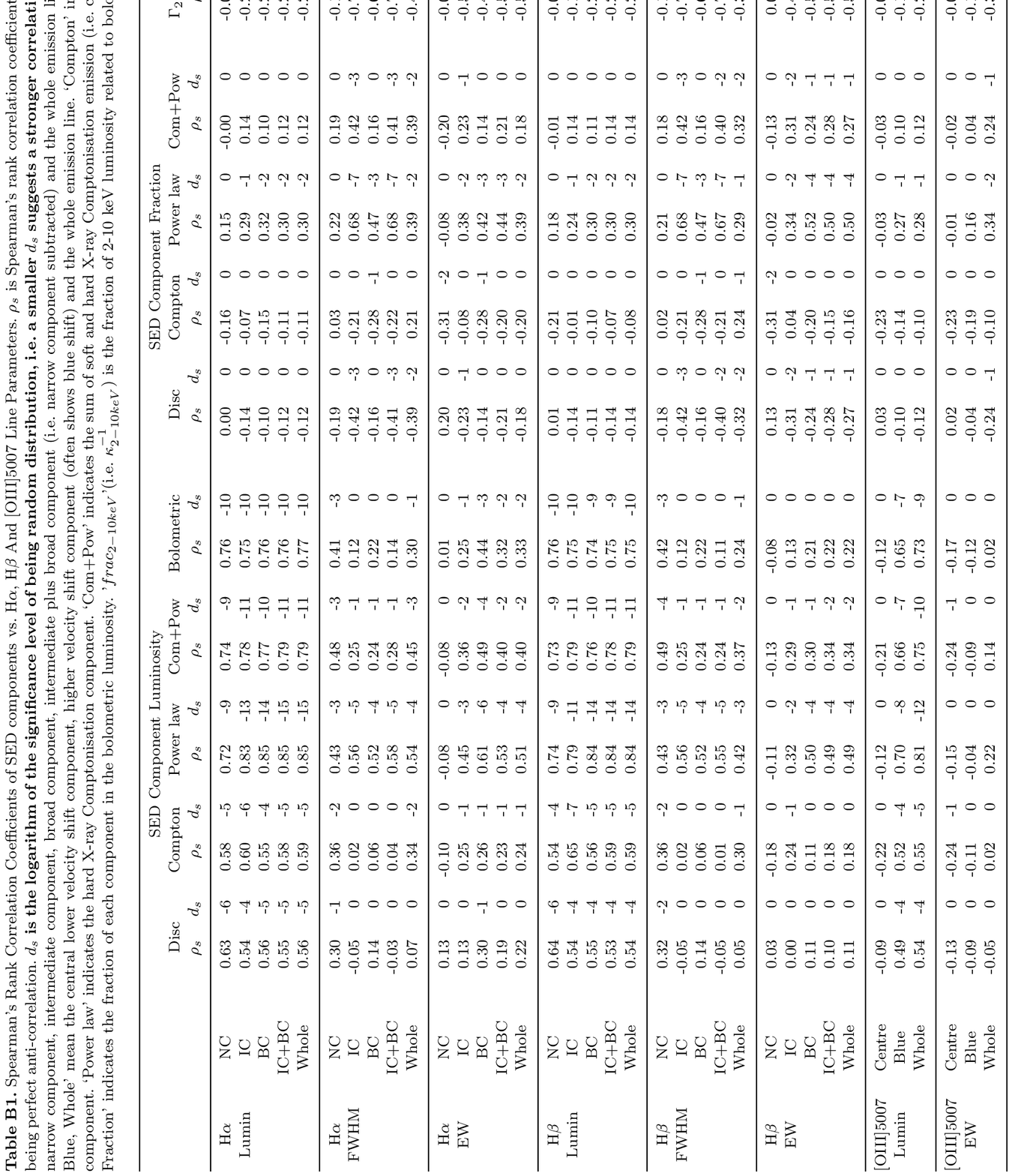}
        \end{tabular}
    \end{minipage}
    \label{app:table:spearman}
\end{table*}

\end{document}